\documentclass{pasj00}

\begin{document}
\SetRunningHead{Author(s) in page-head}
{Origins of Dispersions of $E_{p}$--Brightness Correlation in GRBs}

\title{Possible Origins of Dispersion of the Peak Energy--Brightness 
Correlations of Gamma-Ray Bursts}

\author{Daisuke \textsc{Yonetoku},\altaffilmark{1}\email{yonetoku@astro.s.kanazawa-u.ac.jp (DY)}
Toshio \textsc{Murakami},\altaffilmark{1}
Ryo \textsc{Tsutsui},\altaffilmark{2}
Takashi \textsc{Nakamura},\altaffilmark{2}
Yoshiyuki \textsc{Morihara}\altaffilmark{1}
and
Keitaro \textsc{Takahashi},\altaffilmark{3}
}
\altaffiltext{1}{Department of Physics, Kanazawa University, Kakuma, Kanazawa, Ishikawa 920-1192, Japan}
\email{yonetoku@astro.s.kanazawa-u.ac.jp (DY)}
\altaffiltext{2}{Department of Physics, Kyoto University,
Kyoto 606-8502, Japan}
\altaffiltext{3}{Department of Physics and Astrophysics,
Nagoya University, Fro-cho, Chikusa-ku, Nagoya, 464-8602, Japan}

%


\maketitle

\begin{abstract}
We collect and reanalyze about 200~GRB data of prompt-emission
with known redshift observed until the end of 2009, and select
101~GRBs which were well observed to have good spectral parameters
to determine the spectral peak energy ($E_p$), 
1-second peak luminosity ($L_p$) and isotropic energy
($E_{\rm iso}$). Using our newly-constructed database with 101 GRBs,
we first revise the $E_p$--$L_p$ and $E_p$--$E_{\rm iso}$ correlations.
The correlation coefficients of the revised correlations are
0.889 for 99 degree of freedom for the $E_p$--$L_p$ correlation and
0.867 for 96 degree of freedom for the $E_p$--$E_{\rm iso}$ correlation.
These values correspond to the chance probability of
$2.18 \times 10^{-35}$ and $4.27 \times 10^{-31}$, respectively.
It is a very important issue whether these tight correlations are 
intrinsic property of GRBs or caused by some selection effect
of observations. In this paper, we examine how the truncation of
the detector sensitivity affects the correlations, and we conclude
they are surely intrinsic properties of GRBs.
Next we investigate origins of the dispersion of the correlations
by studying their brightness and redshift dependence.
Here the brightness (flux or fluence) dependence
would be regarded as an estimator of the bias due to
the detector threshold. We find a weak fluence-dependence
in the $E_p$--$E_{\rm iso}$ correlations and a redshift dependence
in the $E_p$--$L_p$ correlation both with 2~$\sigma$ statistical
level. These two effects may contribute to the dispersion of
the correlations which is larger than the statistical uncertainty.
We discuss a possible reason of these dependence and give
a future prospect to improve the correlations.

\end{abstract}

\section{Introduction \label{sec:introduction}}

There are several correlations between the rest-frame physical
quantities of GRB and their luminosity (or isotropic equivalent
energy $E_{\rm iso}$). The first report was a variability--luminosity
correlation by \citet{fenimore} which states that more variable events
are more luminous. The next one was the lag--luminosity correlation
reported by \citet{norris} and \citet{schaefer01}. Each pulse in 
the prompt emission has a spectral time lag, that is, a time delay 
of the soft-band emission compared with the hard-band one. 
According to this correlation, the events with large spectral time lag 
are dimmer than ones with short time lag. These two correlations are 
based on the temporal behaviors of prompt emission.

Several correlations concerning the spectral property have also
been suggested. \citet{lloyd2000} found a correlation between 
the spectral peak energy ($E_p$) and observed energy fluence.
They also mentioned the possibility of correlation 
in the rest frame of GRBs. \citet{amati02} exactly 
mentioned a very tight correlation between the $E_p$ and 
the isotropic equivalent energy ($E_{\rm iso}$) in the GRB frame
(see also \cite{amati06}). This $E_p$--$E_{\rm iso}$ correlation 
was confirmed and extended toward X-ray flashes by
\citet{sakamoto04} and \citet{lamb04}.
Independently, \citet{yonetoku04} reported similar but
tighter correlation between $E_p$ and the 1-second peak
luminosity ($L_p$) called  $E_p$--$L_p$ correlation.
Moreover, using the GRBs with measured opening half-angle,
\citet{ggl04} found that $E_p$ strongly correlates with
the collimation-corrected gamma-ray energy ($E_{\gamma}$).
\citet{Firmani2006} reported the correlation among 
$E_p$--$E_{\rm iso}$--$T_{0.45}$, where $T_{0.45}$ is the time
spanned by the brightest 45~\% of the total counts above 
the background.

These correlations can be used as cosmological tools to investigate
the physical environment of the early universe.
\citet{yonetoku04} used the $E_p$--$L_p$ correlation as a redshift
indicator for 689 GRB samples observed by BATSE without known redshift, 
and derived the GRB formation rate. Based on this GRB formation rate,
\citet{murakami05} investigated the cosmic reionization epoch
and the metal enrichment by the population-III stars.
On the other hand, they are also useful tools for extending
the Hubble diagram to probe the cosmological expansion history
\citep{takahashi,oguri,ghirlanda06,schaefer07}. \citet{kodama08} 
calibrated the $E_p$--$L_p$ correlation of nearby GRBs with 
the luminosity distance measured by the Type Ia supernovae 
(see also \cite{liang08,cardone09}). They succeeded in
extending the cosmic distance ladder toward the redshift of
$z \sim 6$, and estimated the energy density of dark matter and
dark energy in high-redshift universe beyond $z > 2$. Furthermore,
\citet{tsutsui09} improved the $E_p$--$E_{\rm iso}$ and
the $E_p$--$L_p$ correlation introducing another parameter named 
luminosity time defined as $T_L \equiv E_{\rm iso}/L_p$. 
Using this newly discovered $E_p$--$L_p$--$T_L$ plane, 
they constrained the amount of dark matter and dark energy in 
the early universe more effectively. These papers well demonstrate 
the validity of these empirical correlations as cosmological tools.

However, \citet{np04} gave an argument against the presence of
the $E_p$--$E_{\rm iso}$ correlation. They insisted that about $40\%$
of 751 BATSE GRBs without known redshift do not satisfy 
the $E_p$--$E_{\rm iso}$ correlation even if they assume any redshift, 
and also mentioned that clear outliers such as GRB~980425 and 
GRB~031203 exist. A similar argument against the $E_p$--$E_{\rm iso}$ 
correlation has also been made by \citet{bp05}. They concluded that 
$88$~\% of GRBs detected by BATSE cannot satisfy the 
$E_p$--$E_{\rm iso}$ correlation because the observed $E_p$--fluence 
ratios of these events exceed its maximum value around $z \sim 4$. 
If these arguments are true, we cannot use these correlations as 
cosmological tools. 
\citet{butler}, using the Bayesian approach to estimate
$E^{\rm obs}_p$ for a lot of Swift events, indicated that dim events 
close to the detector sensitivity would make large scatter on 
the $E_p$--$E_{\rm iso}$ and $E_p$--$L_p$ correlations, and that 
there is a significant threshold effect (see also \cite{lloyd2000}).

On the other hand, \citet{ggf05} gave a positive argument on
the presence of the correlations. They used 442 bright BATSE GRBs
with the pseudo redshifts derived from the lag-luminosity correlation
\citep{bnb04} and obtained the  $E_p$--$E_{\rm iso}$ correlation
with the slightly different power-law index and the larger scatter
than the original one. They found that the chance probability of
the revised correlation is $2.1 \times 10^{-65}$. Similar conclusion
has also been derived by \citet{bclb05}. \citet{ggfcb05} also
checked the validity of the $E_p$--$L_p$ correlation using 442 bright
GRBs with the derived redshift and confirmed the correlation with
the same power-law index within the error in the original
$E_p$--$L_p$ correlation by \citet{yonetoku04}. 
They found that the chance probability of
the $E_p$--$L_p$ correlation is $1.6 \times 10^{-69}$.
The $E_p$--$L_p$ and $E_p$--$E_{\rm iso}$ correlations are 
independently tested by Suzaku-WAM and Fermi-GRB 
\citep{krimm09,amati09}.

These correlations obviously have large dispersion compared with
the statistical fluctuations. The origin of this data scatter is 
unknown, and should be revealed because the reliability and 
the accuracy of the GRB cosmology highly depend on the dispersion 
of correlations. This may be an intrinsic property of GRBs, 
or due to some instrumental threshold effect.
Recently \citet{li07} and \citet{basilakos08} tested the redshift
evolution of the $E_p$--$E_{\rm iso}$, $E_p$--$L_p$,
$E_p$--$E_{\gamma}$ and other correlations
and found no significant evolution, while the statistical
errors are relatively large.
\citet{ghirlanda08}  studied possible instrumental selection
effects on the $E_p$ and Fluence plane in the observer frame. 
In particular, they concentrated on the trigger threshold 
(the minimum peak flux necessary to trigger a given GRB detector)
and the spectral analysis threshold (the minimum fluence necessary 
to determine the value of $E_p$).
They showed these instrumental selection effects do not dominate
for bursts detected before the launch of the $\it Swift$ satellite,
while the spectral analysis threshold may be the dominant truncation
effect of the $\it Swift$ GRB sample.
\citet{nava08} found that  the $E_p$ of the fainter BATSE bursts
is correlated with the fluence and flux with a correlation slope
flatter than the one defined by the known redshift samples.
They showed selection effects are not responsible for
the correlations. About 6~\% of these bursts are surely outliers of 
the $E_p$--$E_{\rm iso}$ correlation, whereas there is only one sure 
outlier on $E_p$--$L_p$ correlation.
 
In this paper, using our newly-constructed database, we investigate
whether the $E_p$--$L_p$ and the $E_p$--$E_{\rm iso}$ correlations
represent the intrinsic property of GRBs or artifacts of
the truncation effect due to the detector sensitivity.
Then, we study the origin of the dispersion of the correlations.
In particular, we consider two possible origins, threshold effect
and redshift evolution. The former is that samples which are
detected marginally above the detector threshold may cause
some systematic error. This should be tested by dividing
the samples according to how larger than the threshold
the flux/fluence is. However, the threshold here should not
be the naive one because we are focusing on samples with
the spectral parameters. Thus we have to use so called spectral
threshold \citep{nava08} which needs detailed simulations.
Instead, we simply study flux/fluence dependence of the correlations
as an estimator for the systematic error due to the threshold effect.
This would be justified because the spectral thresholds
for many detectors are roughly the same as shown in \citet{nava08}.

The structure of this paper is as follows. First we present
a database of 101~GRBs with known redshift and well-determined
spectral parameters observed by the several independent missions
in section~\ref{sec:database}. This database will be useful to
estimate the intrinsic property of GRBs and we update
the $E_p$--$L_p$ and $E_p$--$E_{\rm iso}$ correlations in section
\ref{sec:correlation}. In section~\ref{sec:selection}, we examine
the possible systematic effects of these correlations such as
the dependence of the correlations on the flux and fluence,
and the redshift evolution. Finally we will give discussion and
summary in section~\ref{sec:discussion}.

\section{Data Selections and Analyses} \label{sec:database}

We collected and reanalysed about 200 GRBs obtained by several 
instruments aboard the independent missions. Each instrument covers 
a different energy range with a different time resolution, 
so we should treat them carefully when we compare physical 
quantities such as $E_p$ and $L_p$. In this section, 
we gather the observational data obtained until the end of 2009 
and derive physical quantities by uniform criteria to construct 
a reliable database. After that, using this database, we discuss 
the intrinsic property of the prompt emission of GRBs.

The prompt gamma-ray spectrum can be usually described as
the spectral model of the exponentially-connected broken power-law
function suggested by \citet{band93}:
\begin{eqnarray}\label{eq:band}
N(E) = \left\{
\begin{array}{ll}
        A \Bigl( \frac{E}{100~{\rm keV}} \Bigr)^{\alpha}
        \exp(- \frac{E}{E_{0}})
        &
        {\rm for} \ E \le (\alpha - \beta) E_{0},\\
        A \Bigl( \frac{E}{100~{\rm keV}} \Bigr)^{\beta} \Bigl(
        \frac{(\alpha - \beta) E_{0}}{100~{\rm keV}}\Bigr)^{\alpha - \beta}
        \exp(\beta - \alpha)
        &
        {\rm for} \ E > (\alpha - \beta) E_{0}.
\end{array}
\right.
\end{eqnarray}
Here $N(E)$ is in units of ${\rm photons~cm^{-2} s^{-1} keV^{-1}}$.
This function has four parameters, the low-energy photon index $\alpha$,
the high-energy photon index $\beta$, the spectral break energy $E_0$
and the normalization $A$. The peak energy can be derived as
$E^{\rm obs}_p = (2+\alpha) E_{0}$, which corresponds to the energy
at the maximum flux in the $\nu F_{\nu}$ spectra.
In this paper, we simply denote that $E_{p} = E^{\rm obs}_{p}(1+z)$
in the rest frame of GRB.

According to BATSE observations, the average properties of 
low- and high-energy spectral indices are $\alpha \sim -1$ 
and $\beta \sim -2.25$, respectively \citep{preece98, preece00}.
The recent Fermi-LAT observations support that the high energy
power-law index is consistent with the typical value of
$\beta = -2.25$ beyond 1~GeV energy range
(e.g. GRB~080916C, 081024B, 090323, 090428).
Therefore, in this paper, if the high energy index has not been 
measured, we assume $\beta = -2.25$ as a fixed value
when we estimate the bolometric flux and fluence.

Here it should be noted that the spectra of some GRBs are well
fitted by cutoff power-law model. If some events show in fact 
the cutoff spectra, we will overestimate their bolometric flux 
and fluence when we fit the spectrum with the Band function 
\citep{shahmoradi09}. However, as far as we know, there is almost 
no positive evidence about the existence of a clear cutoff 
in the prompt gamma-ray spectrum. Although many GRB spectra have 
been fitted with the cutoff power-law model, most of them are 
also well fitted by the Band function
(see \cite{kaneko06} and \cite{gcn8256} (GRB~080913), 
\cite{gcn9422} (GRB~090516) and \cite{gcn9821} (GRB~090812) 
for more recent events). 
Therefore, at present, the fitting results by cutoff power-law model 
may tend to underestimate their total fluence and flux. 
Furthermore, as shown in \citet{kaneko06}, it often happens that
$\beta$ cannot be determined by the data and the cutoff
power-law model is sufficient to fit, if the peak energy is
close to the high-energy end of the detector sensitivity,
or if the event is so dim that the number of high energy photons
is very small. Interestingly, simulations in \citet{kaneko06}
showed that, if the signal-to-noise ratio is relatively low,
a spectrum with the shape of the Band function is well fitted
by cutoff power-law model. Thus, the assumption of the Band
function for all events including dimmer GRBs would be reasonable 
at present. Our database is uniformly constructed in this sense.

For the purpose of the complete coverage of database,
we adopt three event selection criteria.
(1) We selected GRBs with  known redshift observed until the end
of 2009. We have about 200 samples under this criterion from GCN
Circular Archive \citep{gcn} and GRBlog \citep{quimby03}.
(2) We selected samples whose spectral parameters are well measured.
(3) We used samples whose total fluence and 1-second peak flux
are reported.
Under these three criteria we have 101 GRBs. For these samples,
we can estimate the bolometric energy and the peak luminosity by
extending the spectral parameters toward the appropriate energy band
for each GRB with known redshift. In this paper, referring
the previous works, we use 1--10,000~keV energy band in the rest frame
of GRBs when we calculate the bolometric energy and peak luminosity.
Then, we have to convert the observed fluence ($S_{\rm obs}$) and
the 1-second peak photon-flux ($P_{\rm p,obs}$) within the energy
range between $E_{\rm min}$ and $E_{\rm max}$ of each instrument,
into the bolometric fluence ($S_{\rm bol}$) and
the bolometric 1-second peak energy-flux ($F_{\rm p,bol}$) as
\begin{eqnarray}
\label{eq:conversion}
S_{\rm bol} &=& S_{\rm obs} \times
\frac{\int_{1/(1+z)}^{10,000/(1+z)} E \times N(E) dE}
{\int_{E_{\rm min}}^{E_{\rm max}} E \times N(E) dE}~~~{\rm (erg~cm^{-2})},\\
\label{eq:conversion2}
F_{\rm p,bol} &=& P_{\rm p,obs} \times
\frac{\int_{1/(1+z)}^{10,000/(1+z)} E \times N(E) dE}
     {\int_{E_{\rm min}}^{E_{\rm max}} N(E) dE}~~~{\rm (erg~cm^{-2}s^{-1})}.
\end{eqnarray}
Here, the integration is performed between the energy range of
$1/(1+z)~{\rm keV}$ and $10,000/(1+z)~{\rm keV}$ for the Band
function $N(E)$. This is equivalent to the $k$-correction.
Then the bolometric isotropic energy ($E_{\rm iso}$)
and the 1-second peak luminosity ($L_p$) of 1--10,000~keV
in the rest frame of GRB can be simply calculated as
\begin{eqnarray}
\label{eq:eiso}
E_{\rm iso} &=& \frac{4 \pi d_L^2 S_{\rm bol}}{1+z}~~~{\rm (erg)},\\
\label{eq:lp}
L_p &=& 4 \pi d_L^2 F_{\rm p,bol}~~~{\rm (erg~s^{-1})}.
\end{eqnarray}
Here, $d_L$ is the luminosity distance calculated with
the cosmological parameters of
$(\Omega_{\rm m}, \Omega_{\Lambda}) = (0.3, 0.7)$ and the Hubble
parameter of $H_0 = 70~{\rm km~s^{-1} Mpc^{-1}}$.

It should be noted that we have very small number of GRBs 
satisfied with all three selection criteria if we use 
the information observed by a single instrument for each GRB. 
For example, the Swift/BAT is a powerful instrument to measure 
the low energy photon index $\alpha$ while it is difficult to 
determine the high energy index $\beta$, and the $E^{\rm obs}_p$ 
value is not available for most GRBs. However the hard X-ray 
instruments, e.g. Konus, HXD-WAM and RHESSI strongly support 
Swift/BAT by measuring the higher-energy part of prompt spectra.
Thus we frequently merge the spectral information reported by
independent missions to obtain the entire shape of the spectrum.

However, there are some GRBs observed by the independent satellites
whose spectral parameters are inconsistent with each other.
In these cases, we basically exclude these samples from the table
because we can not find any reasonable reason to choose one.
However, there are two exceptions, GRB090423 and GRB090424.
Fermi-GBM has a great advantage for the spectral measurement,
especially for the $E^{\rm obs}_p$ determination, thanks to
its wide band energy coverage. We chose the Fermi-GBM data rather
than the Swift-BAT data for the two events, which have
inconsistent $E^{\rm obs}_p$ values by the two detectors.
Here we would like to emphasize that $E^{\rm obs}_p$ values
given by independent detectors are consistent with each other
for a large majority of events within 1~$\sigma$ error.
In those cases, we chose the one with a smaller error for them.

\begin{figure}
\begin{center}
\rotatebox{270}{\includegraphics[width=60mm]{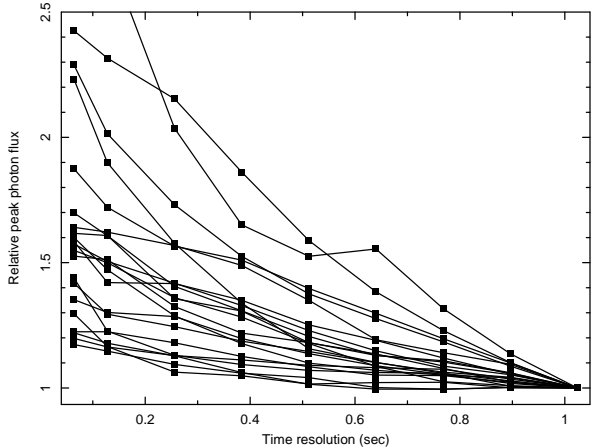}}
\end{center}
\caption{Relative peak photon fluxes for different time interval
normalized by the one at 1024~msec. We used the lightcurve data
with 64~msec time resolution observed by Konus. Peak photon flux
with shorter time interval is systematically brighter than 
the 1024~msec one. In this paper, we correct the time scale of 
peak fluxes if the reported value is in different time scale such as
64~msec or 256~msec.}
\label{peak-flux}
\end{figure}

Almost all satellite teams report the fluence ($S_{\rm obs}$) and
1-second peak photon flux ($P_{\rm p,obs}$) in the energy range of
their own instruments. However, some Konus reports did not include
the 1-second peak flux but only 64~msec or 256~msec one.
Then using the lightcurve data published by the Konus team,
we reanalyzed them and estimated the 1~second peak flux.
The typical time resolution of the Konus/Wind lightcurve is
64~msec. We performed the re-binning of 16 channel time bins
into 1024~msec one with the running average method.
Although we have 16 degrees of freedom (d.o.f) to choose the start point
when we calculate the 1~second peak flux, here, the 1~second peak flux 
is determined as the brightest one. This is a quantitatively important
modification which has been missed in the previous works.
In figure~\ref{peak-flux}, we show relative peak photon fluxes of
Konus events for different time intervals (64, 128 ,256, 384, 512,
640, 768, 896 and 1024~msec) normalized by 1024~msec peak flux
for each event. It can be seen that 
the 64~msec peak flux is systematically brighter than 
the 1~second one by  60--70~\% level on average.

\begin{figure}
\rotatebox{0}{\includegraphics[width=80mm]{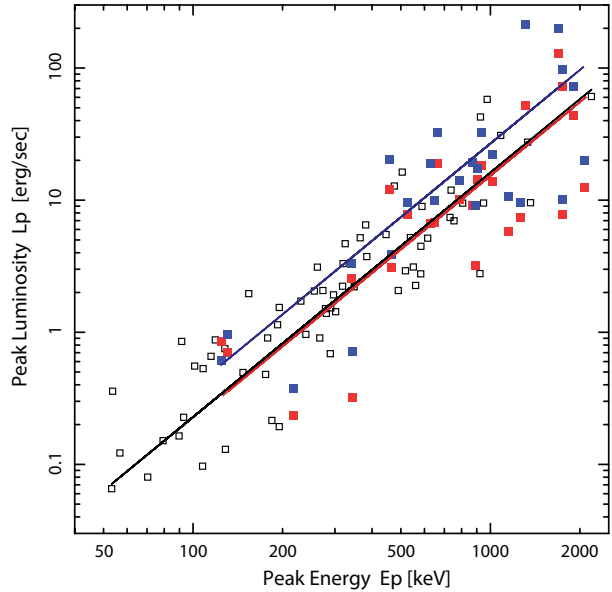}}
\caption{The $E_{p}$--$L_{p}$ distributions of Konus data.
The blue and the red solid squares are the peak luminosity measured with 
64~msec and 1024~msec time scale in the observer frame, respectively. 
The black open squares are the data 1-second (or 1024~msec) peak luminosity 
observed by the other missions. The black, red and blue solid lines 
are the best fit power-law functions for the same colour data with 
the fixed index, respectively. The blue data systematically distribute 
higher than the black line with the factor of 1.70. On the other hand, 
the red points are consistent with the black line.}
\label{Konus-1024ms}
\end{figure}

In figure~\ref{Konus-1024ms}, we show the $E_{p}$--$L_{p}$ distributions
of the Konus data. The blue and the red filled squares are 
the peak luminosity measured with 64~msec and 1024~msec time scale 
in the observer frame, respectively. The black open squares are 
the data of 1~second peak luminosity observed by the other missions.
The black, red and blue solid lines are the best fit power-law 
functions for the same colour data with the fixed index, respectively. 
We can find that the blue solid line is apart from the other two lines,
and the normalization of the blue line is 1.70 times higher than the
black one. On the other hand, the red line is almost same as 
the black one with the factor of 0.96. This fact indicates that 
the 64~msec peak luminosity makes systematic dispersion in 
the $E_{p}$--$L_{p}$ correlation. 
Several past results, i.e. \citet{nava08, ghirlanda09} as well as
our results by \citet{kodama08, tsutsui09}, were argued about 
the selection effect and the GRB cosmology using the 64~msec peak 
luminosity for the Konus data. At present, since the number fraction 
of the Konus data is about 25~\% of the entire 101 samples, 
the definition of the time scale should be treated correctly.
Therefore the newly constructed database in this paper is 
more uniform compared with the past database, and appropriate 
to examine the origin of the data dispersion on the $E_{p}$--$L_{p}$ 
correlation.

The data are summarized in table~\ref{tab:data1} in
appendix~\ref{sec:appendix}. Additionally, we also show 
the data excluded from the database in table~\ref{tab:excludedata1} 
and \ref{tab:excludedata2} because of several reasons. 
For example, we know that so-called low luminosity GRBs and outliers are 
really exist. In this paper, we call the data ``outlier'' which locates 
outside of the 3~$\sigma$ confidence region of the $E_{p}$--$L_{p}$ and
the $E_{p}$--$E_{iso}$ correlations (see section~\ref{sec:correlation}). 
These groups are summarized in table~\ref{tab:excludedata1}.
In table~\ref{tab:excludedata2}, we show several events with large 
ambiguity on the redshift and the spectral parameters. 
Several famous short GRBs are also listed in the same table as 
a reference, since their peak luminosity is determined by 
the millisecond time scale, 
and we can not convert them into the 1-second peak luminosity. 
In the following sections~\ref{sec:correlation}, 
we discuss the $E_p$--$L_p$ and $E_p$--$E_{\rm iso}$ correlation.
After that, in section~\ref{sec:selection}, we examine a flux dependence
and an redshift evolution effect for these two correlations with
table~\ref{tab:data1}.

\section{Correlations} \label{sec:correlation}

\begin{figure}
\rotatebox{0}{\includegraphics[width=100mm]{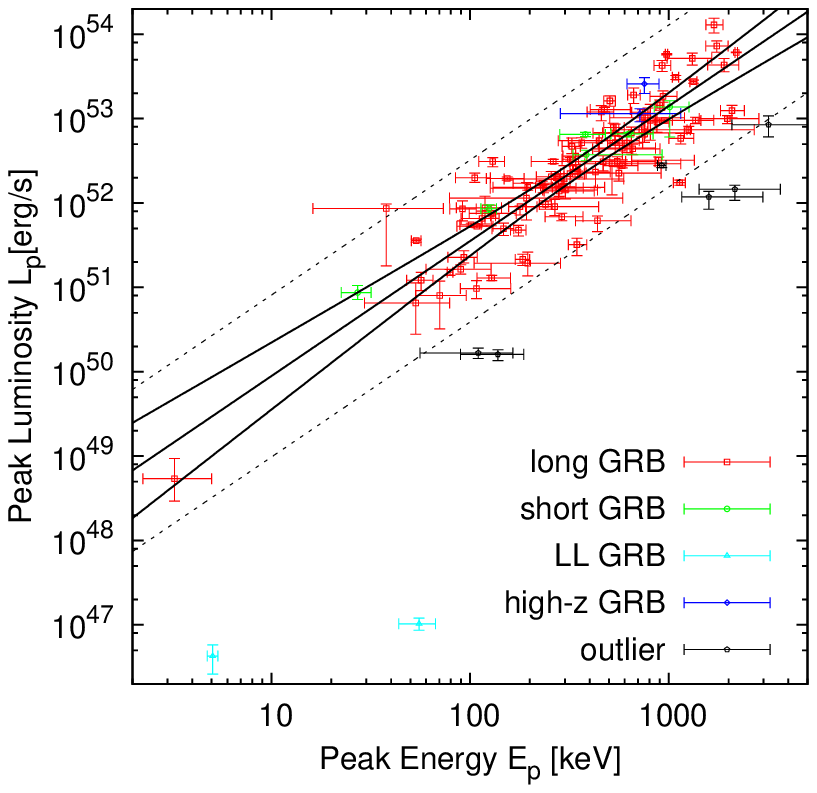}}
\rotatebox{0}{\includegraphics[width=100mm]{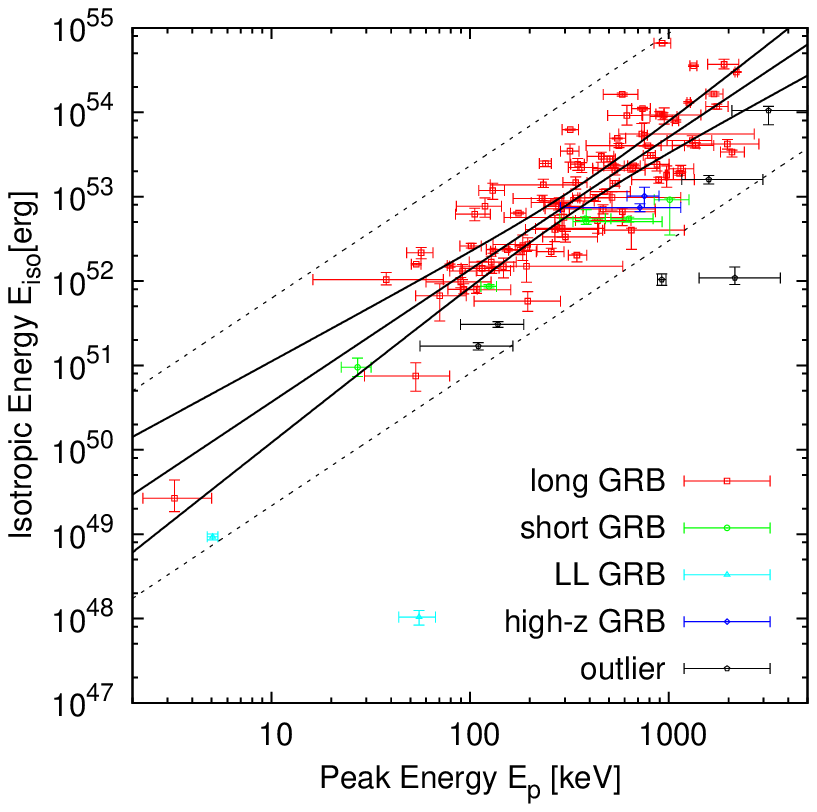}}
 \caption{The $E_p$--$L_p$ correlation (top) and
the $E_p$--$E_{\rm iso}$ correlation (bottom) for all events 
listed in table~\ref{tab:data1} and \ref{tab:excludedata1}. 
The red and green points indicate the data of long-GRBs and 
short-GRBs defined by \citet{levesque2010} in the rest frame of GRBs. 
The 2~light-blue plots are the well-known low-luminosity GRBs of 
GRB~980425 and GRB~060218. The blue points are high-redshift GRBs 
with $z>6$ (GRB~080913, GRB~090423). 
The solid black straight lines are the best fit function for 
long (red) and short (green) GRBs while two black curves around 
the straight line are 3~$\sigma$ statistical error. 
The dotted lines are 3~$\sigma$ systematic errors in 
equations~\ref{Eq:yonetoku} and \ref{Eq:amati}.
The 6 black plots are outliers which locates beyond 3~$\sigma$ 
confidence region from the best fit function of the $E_p$--$L_p$ 
and/or the $E_p$--$E_{\rm iso}$ correlations
(GRB~050223, 050803, 050904, 070714B, 090418, and 091003).}
\label{Ep-Lp-Eiso}
\end{figure}

In figure~\ref{Ep-Lp-Eiso}, we show the $E_p$--$L_p$ (upper) and 
the $E_p$--$E_{\rm iso}$ (lower) correlations for all events listed 
in table~\ref{tab:data1} and table~\ref{tab:excludedata1}. 
We used the equations~\ref{eq:eiso} and \ref{eq:lp}
when we estimate the isotropic energy and the 1-second peak 
luminosity, respectively. The solid black straight lines are
the best-fit power-law functions for good data listed in table~\ref{tab:data1}.
The black curves around the straight lines represent 3~$\sigma$ 
statistical errors defined as,
\begin{equation}
\sigma_{\log{L_p (E_{\rm iso})}}
= \sqrt{\sigma_A^2
         + \left( \sigma_B \log \left(\frac{E_p}{\rm 355~keV}\right)
  \right)^{2}}, 
\end{equation} 
where $\sigma_A$ and $\sigma_B$ are errors on the normalization 
and the power-law index when we express the correlation as 
$L_{p}({\rm or}~E_{iso}) = A E_{p}^{B}$, respectively. 
The dotted lines are 3~$\sigma$ confidence regions including the data 
dispersion (systematic error).

In order to reduce the effect of samples which have unexpectedly
large systematic error or which constitute different (unknown)
family of GRBs, we have to identify and remove outliers.
We identified the outliers as follows. First, we use all samples
except the two low-luminosity GRBs and obtain a tentative
best-fit relation. Then samples which are more than 3-sigma
away from the tentative relation are removed and a new (and still
tentative) best-fit relation is obtained. Performing these
procedures iteratively, we finally obtain the true best-fit relation
and outliers. Thus, the selection of the outliers is not arbitrary.
As a result, we found 6~outliers, represented by the black points
in figure~\ref{Ep-Lp-Eiso}.

Our method is similar to those adopted in the analyses
of Type Ia SNe (e.g. \cite{kowalski2008}) and Cepheid variables
(e.g. \cite{riess2009}). It may be instructive to note that
the number fractions of outliers are about the same for GRBs,
Type Ia SNe and Cepheid.

In figure~\ref{Ep-Lp-Eiso}, each color means the long GRBs (red points), 
high-redshift GRBs ($z > 6$, blue points) and short GRBs (green points),
respectively. Recently, \citet{levesque2010} suggested a definition of
short-hard/long-soft GRBs with a statistical method.
Their definition is roughly
(1) events with $T^{\rm obs}_{90}/(1+z) \le 1.0$~sec are 
short-hard GRBs. 
(2) events with $T^{\rm obs}_{90}/(1+z) \ge 2.0$~sec are 
almost long-soft GRBs. 
(3) events with $1.0 < T^{\rm obs}_{90}/(1+z) < 2.0$~sec 
can not be clearly defined but events with 
$E_p^{\rm obs} > 100~{\rm keV}$ and shorter time scale 
(close to $\sim 1~{\rm sec}$) are likely to be short-hard GRBs.
Here, $T^{\rm obs}_{90}$ is measured as the time duration 
during the 90~\% of total observed photons have been detected while 
the $T^{\rm obs}_{90}$ values highly depend on the energy 
band and the sensitivity of instruments.
In this paper, we refer to their criteria while the original bimodal 
distribution of $T^{\rm obs}_{90}$ measured by BATSE \citep{fishman94} 
is seen in the observer frame. 

We estimated the best fit power-law function 
for the data listed in table~\ref{tab:data1}:
\begin{equation}
L_p = 10^{52.43 \pm 0.037} \times
      \left[ \frac{E_p (1+z)}{355 {\rm keV}} \right]^{1.60\pm 0.082},
\label{Eq:yonetoku}
\end{equation}
\begin{equation}
E_{\rm iso}
= 10^{53.00 \pm 0.045} \times
  \left[ \frac{E_p (1+z)}{355 {\rm keV}} \right]^{1.57 \pm 0.099}.
\label{Eq:amati}
\end{equation}
Here we included not only the statistical errors but also 
the data dispersion (systematic error) with weighting factor 
when we estimated the best fit values and errors in these correlations. 
The data dispersions are estimated as
$\sigma_{{\rm sys},\log L_p}=0.33$ for the $E_p$--$L_p$ correlation and
$\sigma_{{\rm sys},\log E_{\rm iso}}=0.37$ for the $E_p$--$E_{\rm iso}$
correlation, respectively. Then, the correlation coefficients are
{0.889} for 99 d.o.f for the $E_p$--$L_p$ correlation and
{0.867} for 96 d.o.f for the $E_p$--$E_{\rm iso}$ correlation.
These values correspond to the chance probability of
{$2.18 \times 10^{-35}$} and 
{$4.27 \times 10^{-31}$}, respectively.

The functional forms of these two correlations originally proposed
in \citet{yonetoku04} and \citet{amati06} are
\begin{eqnarray}
\frac{L_p}{10^{52}~{\rm erg~s^{-1}}} 
&=& (2.34^{+2.29}_{-1.76}) \times 10^{-5}
\Bigl[ \frac{E_p}{1~{\rm keV}} \Bigr]^{2.0 \pm 0.2},\\
\frac{E_p}{1~{\rm keV}} 
&=& (81 \pm 2) 
\Bigl[\frac{E_{\rm iso}}{10^{52}~{\rm erg}} \Bigr]^{0.57 \pm 0.02},
\end{eqnarray}
respectively. The power-law index of the revised $E_p$--$L_p$
correlation is slightly different from the original one, but 
consistent within 2~$\sigma$ confidence level. 
Considering that \citet{yonetoku04} were able to use only 16 samples, 
2~$\sigma$ level agreement would not be strange.
For the $E_p$--$E_{\rm iso}$ correlation, the power-law index of
our result is consistent with the original one by \citet{amati06}
within 1~$\sigma$ confidence level.

It is a hot topic whether the short GRBs are 
consistent with these correlations or not \citep{ghirlanda09}.
In the criteria of short GRBs by \citet{levesque2010}, 
they are consistent with both correlations within 
the systematic error. However the number of short GRBs is still poor,
so more future observations and arguments should be required.

\section{Possible Origin of Data Dispersion} \label{sec:selection}

As shown in the previous section, the two correlations are
very tight, although there are some dispersion around the best-fit
correlation: $\sigma_{{\rm sys},\log L_p} = 0.33$ and 
$\sigma_{{\rm sys},\log E_{\rm iso}} = 0.37$, respectively.
These correspond to the errors of about factor 2 when we estimate
$L_p$ and $E_{\rm iso}$ from $E_p$.

\begin{figure}
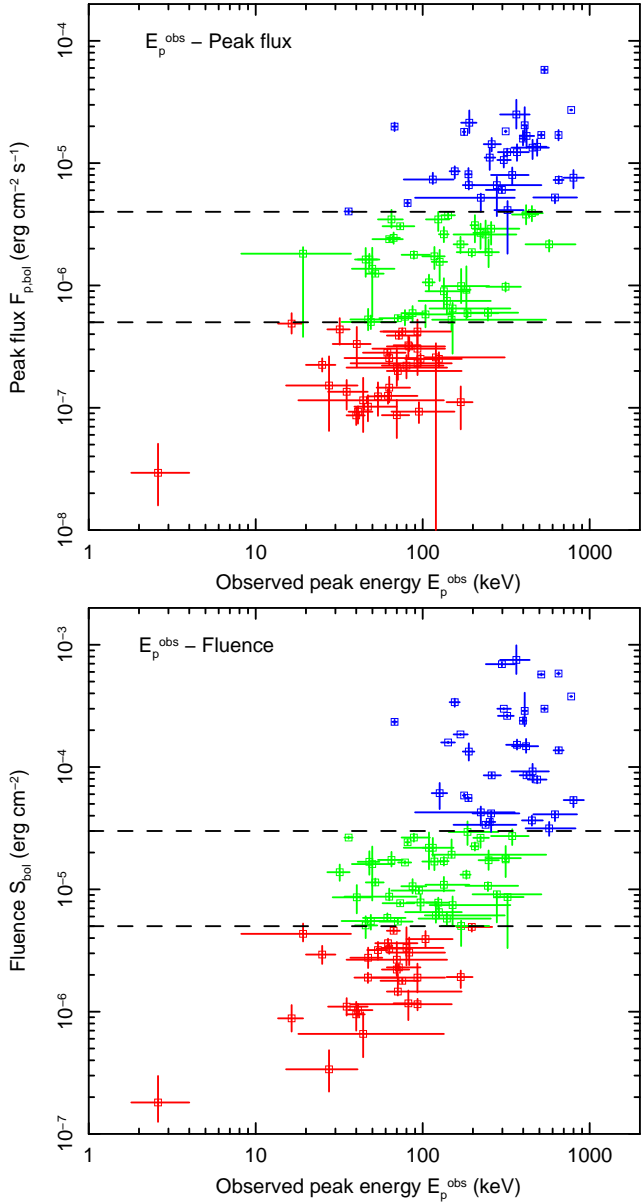

\rotatebox{270}{\includegraphics[width=80mm]{fig4a.ps}}
\rotatebox{270}{\includegraphics[width=80mm]{fig4b.ps}}
\caption{The observed $E_p$--$F_{p, {\rm bol}}$ (left) and 
the $E_p$--$S_{\rm bol}$ (right) correlations. We classified 
the entire data into three groups according to each brightness 
range as listed in table~\ref{table-flux}. 
Each three group does not show the strong correlation 
between $E_{p}$ and brightness, while the entire data set shows
the clear correlation because of the truncation effect by 
the detector threshold. 
If each group consists of each $E_p$--$L_p$ and 
$E_p$--$E_{\rm iso}$ relation, then we may conclude the both 
correlations are real, and not due to the truncation effect.}
\label{Ep-Fp-Sbol}
\end{figure}

One may wonder if these correlations represent the intrinsic property
of GRBs or just come from the truncation effect of the detector
threshold \citep{butler, shahmoradi09}. In figure~\ref{Ep-Fp-Sbol},
we show the sample distributions in observer-frame quantities
($E_p^{\rm obs}$--$F_{p,{\rm bol}}$ plane and
$E_p^{\rm obs}$--$S_{\rm bol}$ plane). We can see some correlations
over three orders of magnitude in the observed brightness range,
but these are substantially affected by the truncation of samples
due to the detector sensitivity. In particular, GRBs with low
brightness and large $E_p^{\rm obs}$ are generally hard to observe.
Some authors insist that the truncation effect create the apparent
correlations in GRB frame, such as the $E_p$--$L_p$ and
$E_p$--$E_{\rm iso}$ correlations.

However, this is not the case as we show. We divide the GRB events
into three groups according to their brightness as shown in
figure~\ref{Ep-Fp-Sbol} and table~\ref{table-flux} (for the details
of the classification see the next subsection).
As one can see, each group does not show significant correlation
in the observer frame and it is evident that the truncation effect
would be very small within each group. However,
as figure~\ref{Ep-Lp-Eiso-Flux} shows, each group shows
a clear correlation in GRB frame. Thus, we can conclude
that the correlations in GRB frame are not caused by
the truncation effect.

In this section, we seek for the origin of the dispersion
of the correlations. The dispersion may be intrinsic and cannot
be reduced, or some unknown systematic errors may contribute
to the dispersion. Specifically we study the flux/fluence
dependence and redshift evolution of the correlations and
estimate the systematic errors accompanying them.

\subsection{Brightness Dependence} \label{subsec:brightness}
Here we test the brightness dependence of the correlations using
our database listed in table~\ref{tab:data1}.
The database consists of events with various brightness.
The brightest event is GRB~991216 with the peak flux of
$F_{\rm p,bol} = 5.80 \times 10^{-5}~{\rm erg~cm^{-2}s^{-1}}$,
while the dimmest one is GRB~020903 with the peak flux of
$F_{\rm p,bol} = 2.94 \times 10^{-8}~{\rm erg~cm^{-2}s^{-1}}$.
The difference is about 3~orders of magnitude.
If the $E_p$--$L_p$ and the $E_p$--$E_{\rm iso}$ correlations 
are affected by the detector sensitivity as suggested by
\citet{butler}, systematic difference according to the brightness
would be seen in the distribution of events on $(E_p,L_p)$ and
$(E_p,E_{\rm iso})$ planes.

We classify GRB events into three groups according to
the bolometric peak flux to discuss the flux dependence of
the $E_p$--$L_p$ correlation, as shown in table~\ref{table-flux}:
{30} events with 
$F_{\rm p, bol} < 5 \times 10^{-7}$ (dim class),
{41} events with
$5 \times 10^{-7} \le F_{\rm p, bol} \le 4 \times 10^{-6}$
(middle class), and
{30} events with 
$F_{\rm p, bol} > 4 \times 10^{-6}$ (bright class).
We derive $E_p$--$L_p$ correlation for each group and investigate 
whether they are consistent with each other. We perform 
a similar analysis for the $E_p$--$E_{\rm iso}$ correlation 
considering three groups according to the bolometric fluence:
{27} events with 
$S_{\rm bol} < 5 \times 10^{-6}$ (dim class),
{41} events with 
$5 \times 10^{-5} \le S_{\rm bol} \le 3 \times 10^{-4}$ (middle class),
{30} events with 
$S_{\rm bol} > 3 \times 10^{-5}$ (bright class)
as shown in table~\ref{table-flux}. 

\begin{figure}
\rotatebox{0}{\includegraphics[width=70mm]{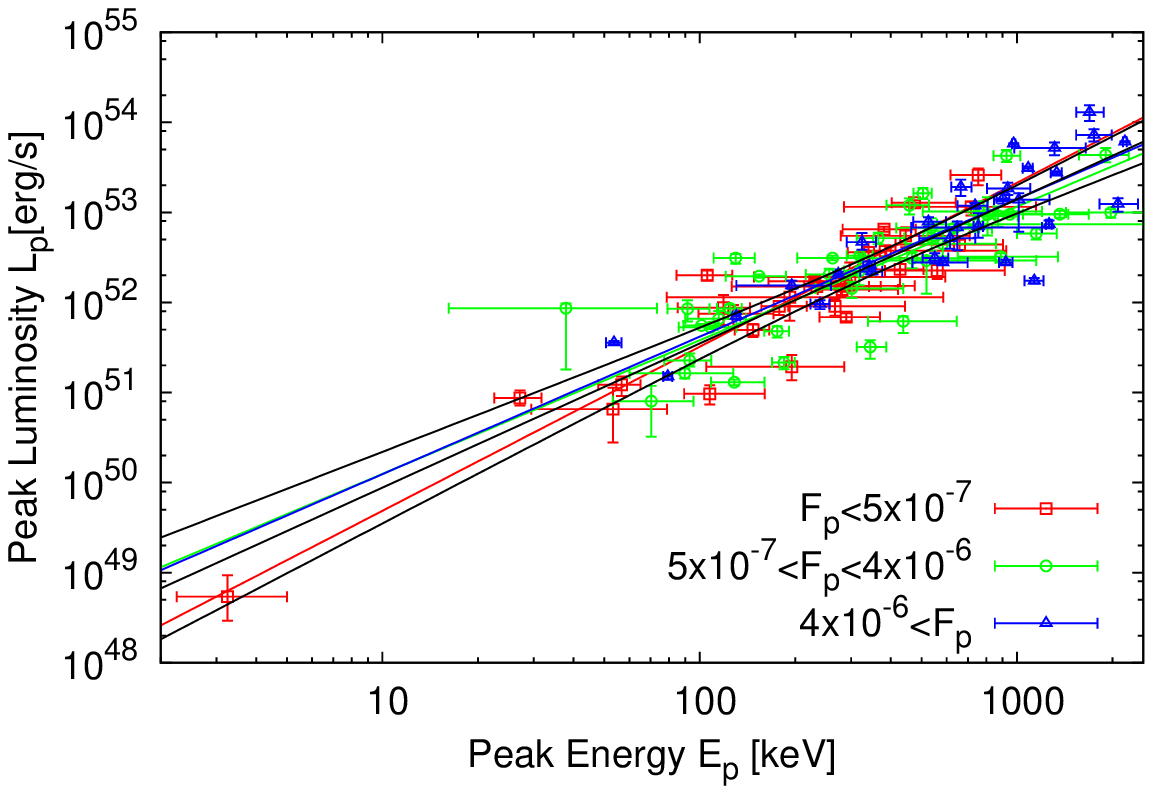}}
\rotatebox{0}{\includegraphics[width=70mm]{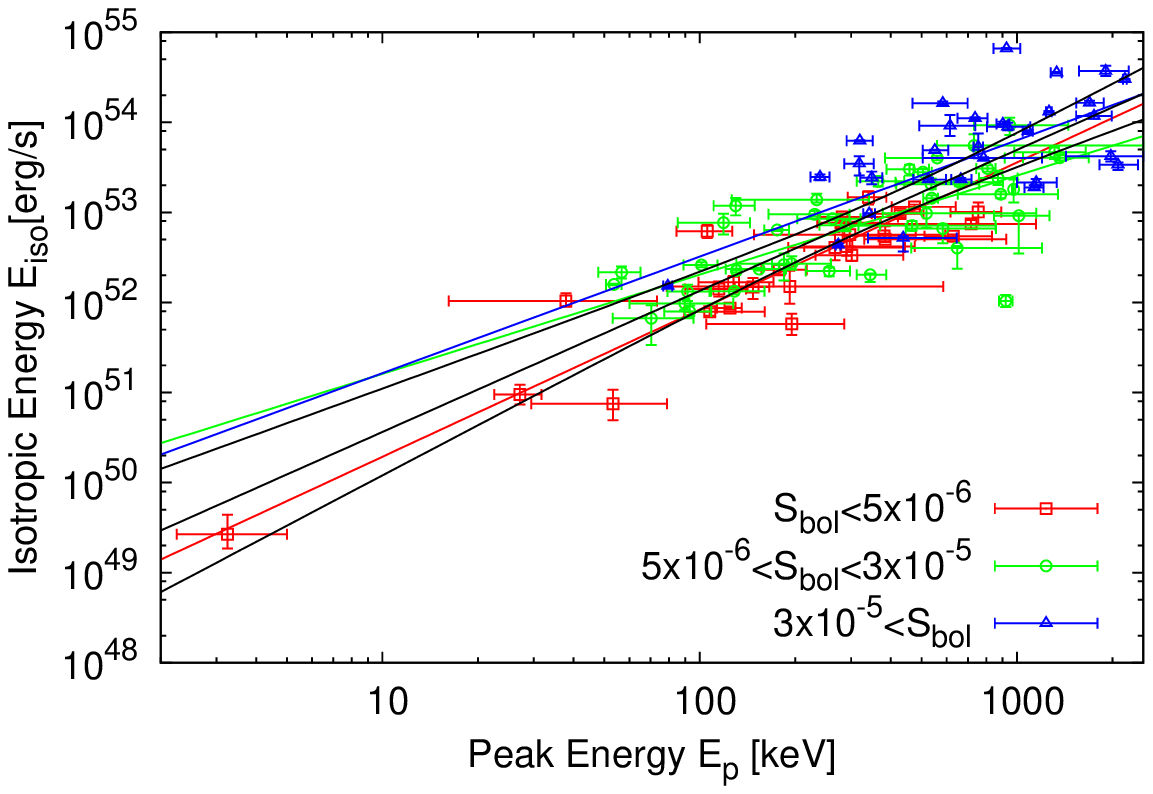}}
\rotatebox{0}{\includegraphics[width=70mm]{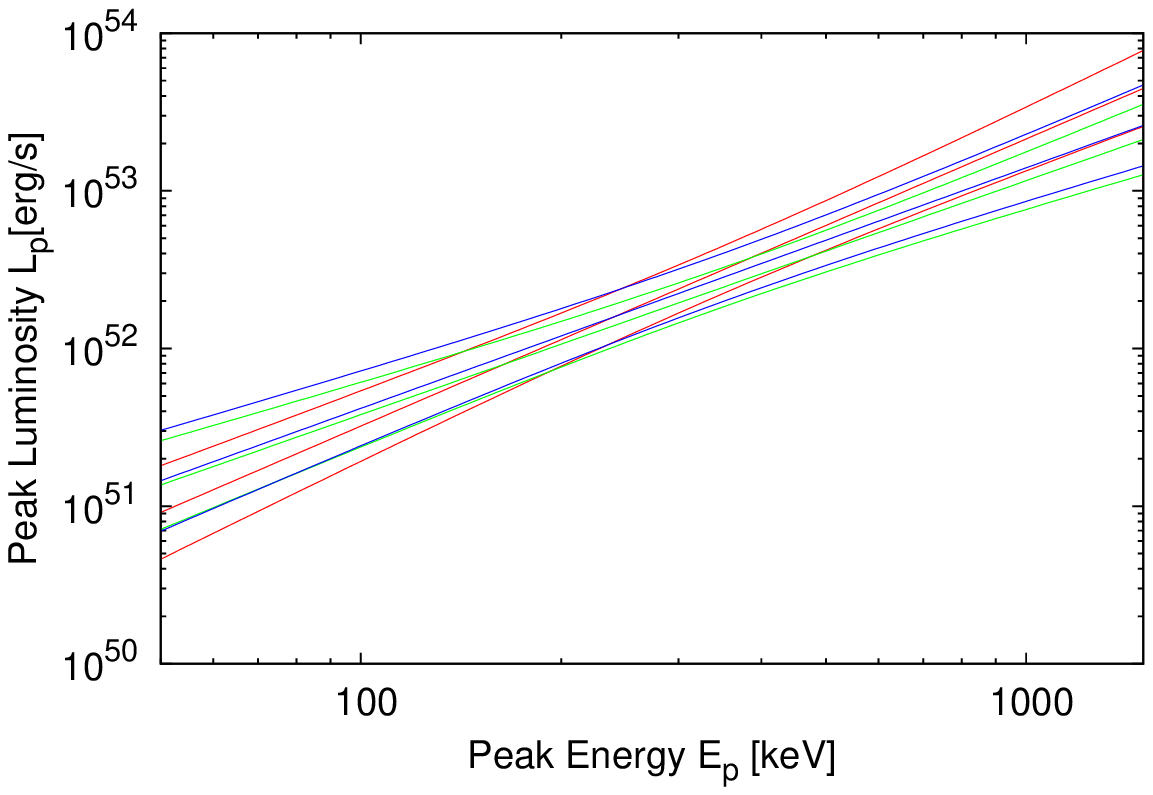}}
\rotatebox{0}{\includegraphics[width=70mm]{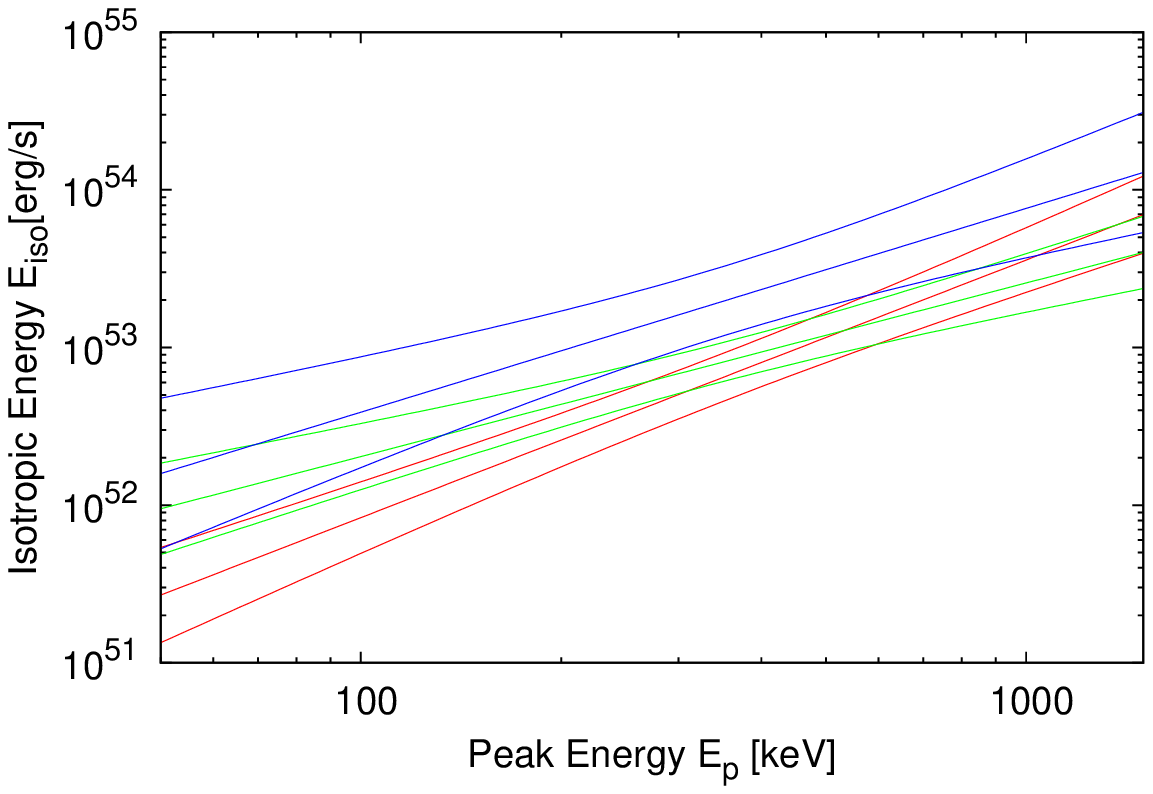}}
\caption{(Top) The left and right panels show the $E_p$--$L_p$ correlation
and the $E_p$--$E_{\rm iso}$ correlation in the three flux (fluence)
ranges listed in table~\ref{table-flux}. Red, green and blue points
represent dim, middle and bright classes, respectively. The black
line and curves are the best-fit function and 3~$\sigma$ statistical
error derived in section \ref{sec:correlation} 
(see equations~\ref{Eq:yonetoku} and \ref{Eq:amati}). 
(Bottom) The left and right
panels show 2~$\sigma$ statistical error regions around the best fit
functions of the $E_p$--$L_p$ and $E_p$--$E_{\rm iso}$ correlations for
the three flux (fluence) ranges listed in table~\ref{table-flux}.
Each colour means the same of top panels. The black line and curves 
are the best-fit function and 2~$\sigma$ statistical error derived 
in section \ref{sec:correlation}. The $E_p$--$L_p$ correlation is
consistent with each other in 1~$\sigma$ statistics while 
the $E_p$--$E_{\rm iso}$ correlation show difference with 2~$\sigma$
confidence level.}
\label{Ep-Lp-Eiso-Flux}
\end{figure}

In figure~\ref{Ep-Lp-Eiso-Flux} (top left and right), 
we show the $E_p$--$L_p$ and the $E_p$--$E_{\rm iso}$ 
correlations for three brightness classes. 
The red, green and blue points represent the dim, middle and
bright class as described in table~\ref{table-flux}, respectively.
We adopt the power-law model for each group, and estimate the best 
fit function as well as the data dispersions of 
$\sigma_{{\rm sys},\log L_p}$ and $\sigma_{{\rm sys},\log E_{\rm iso}}$, 
respectively. The solid line and curves for each colour mean 
the best-fit function and 3~$\sigma$ statistical boundary lines 
as same in figure~\ref{Ep-Lp-Eiso}. 
The results are summarized in table~\ref{fit-flux}.

In figure~\ref{Ep-Lp-Eiso-Flux} (bottom left and right), 
we also show the 2~$\sigma$ acceptable regions for three groups. 
For the $E_p$--$L_p$ correlation, we can recognize that 
three best fit lines are overlapping. So we can say that
the peak-flux dependence is not significant in the $E_p$--$L_p$
correlation. By contrast, for the $E_p$--$E_{\rm iso}$ correlation,
we found a weak trend that the events with larger fluence 
have the larger $E_{\rm iso}$ (or smaller $E_p$).

The discrepancy between the bright and dim classes is over 
2~$\sigma$ statistical level and would contribute to
the dispersion in the $E_p$--$E_{\rm iso}$ correlation.
One should notice that the discrepancy is only a factor of 2 or 3,
while the correlation itself covers beyond 3 or 4 orders of
magnitude in the $E_{\rm iso}$ range. Therefore the detector
sensitivity would contribute to the data dispersion of
the $E_p$--$E_{\rm iso}$ correlation, but not affect the argument
on the existence of the correlation itself. We give an interpretation
of this dependence in section \ref{sec:discussion}.
It should be noted that, as we pointed out in section
\ref{sec:introduction}, the flux/fluence dependence of
the correlations can be regarded as an estimator for
the systematic error due to the threshold effect.


\subsection{Redshift Dependence} \label{subsec:redshift}

\begin{figure}
\rotatebox{0}{\includegraphics[width=70mm]{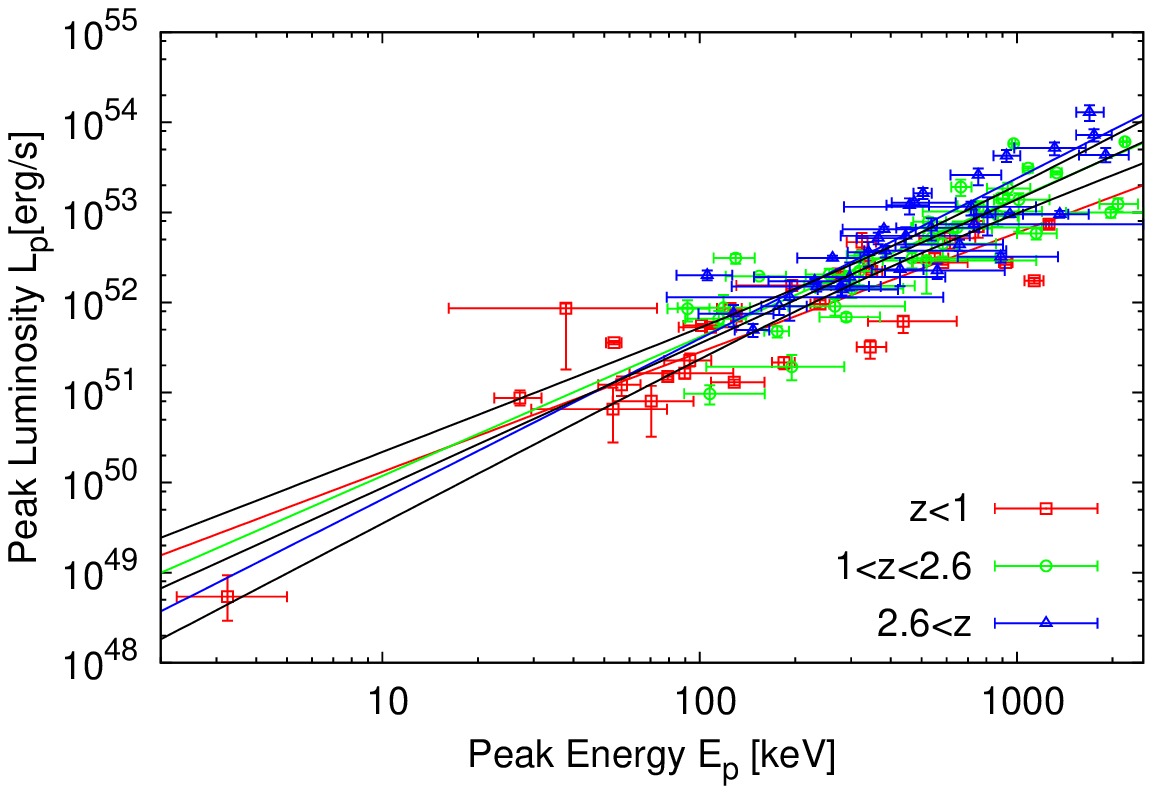}}
\rotatebox{0}{\includegraphics[width=70mm]{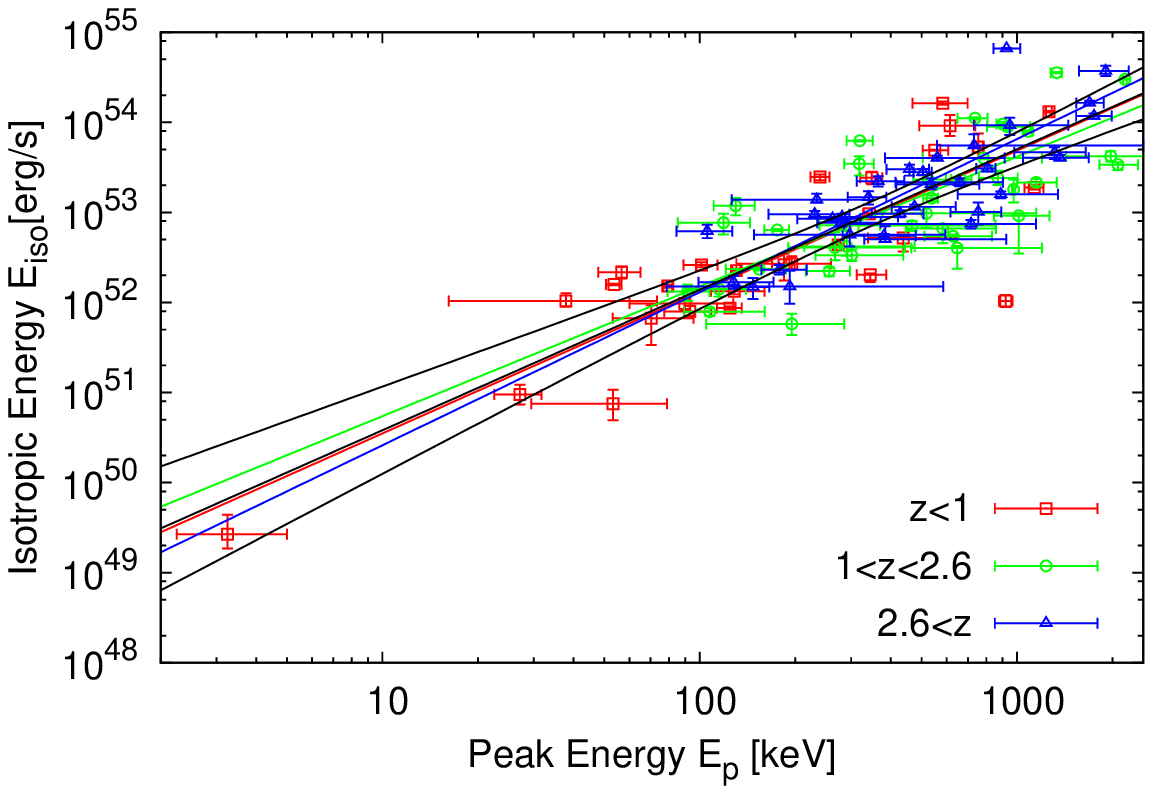}}
\rotatebox{0}{\includegraphics[width=70mm]{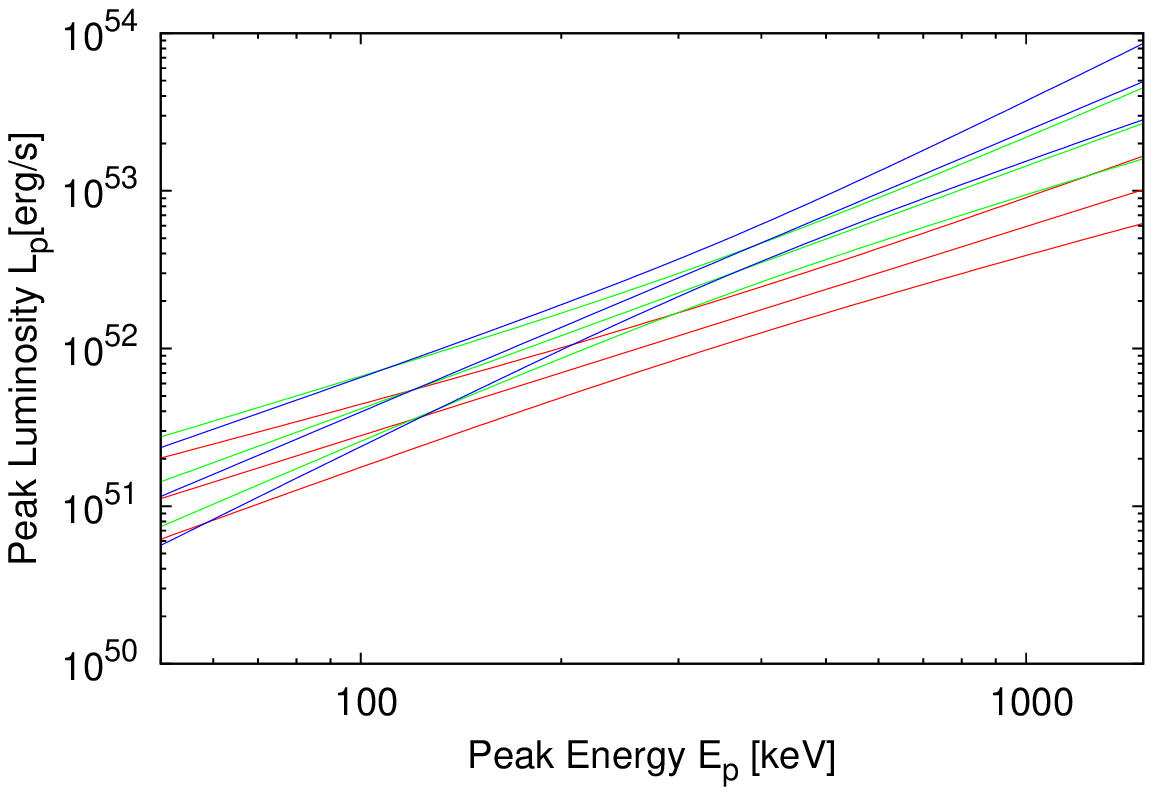}}
\rotatebox{0}{\includegraphics[width=70mm]{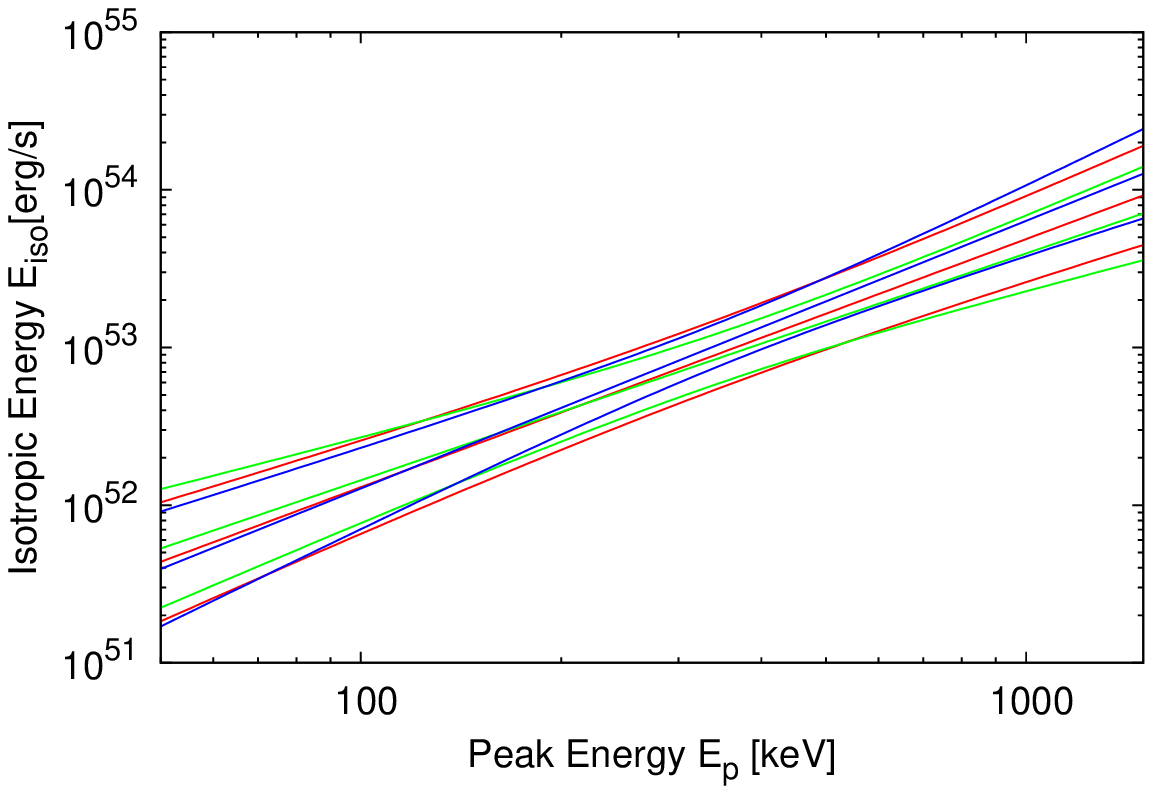}}
\caption{(Top) The left and right panels show the $E_p$--$L_p$
correlation and the $E_p$--$E_{\rm iso}$ correlation in the three
redshift ranges. Each colour represents $z < 1$ (red), 
$1 < z < 2.5$ (green) and $z > 2.5$ (blue), respectively. 
The black line and curves are best-fit function and
3~$\sigma$ statistical error derived in section \ref{sec:correlation}.
(Bottom) The left and right panels show 2~$\sigma$ statistical
error regions around the best-fit functions of the $E_p$--$L_p$
and $E_p$--$E_{\rm iso}$ correlations in the three redshift ranges.
Each colour means the same of top panels.
The black line and curves are best-fit function and 2~$\sigma$
statistical error derived in section \ref{sec:correlation}.
The $E_{p}$--$E_{iso}$ correlation is consistent with each other
in 1~$\sigma$ statistics while the $E_{p}$--$L_{p}$ correlation
show difference with 2~$\sigma$ confidence level.
}
\label{Ep-Lp-Eiso-z}
\end{figure}

We also examine the redshift dependence of the $E_p$--$L_p$ and
the $E_p$--$E_{\rm iso}$ correlations. This is a critical issue when
we use these empirical correlations as cosmological tools like Type Ia
supernovae. Our database covers a wide redshift range of
$0.168 \le z \le 8.2$ and we divide the samples into three classes;
31~GRBs with $z < 1$, 
36~GRBs with $1 < z < 2.5$, and
35~GRBs with $z > 2.5$. 
In figure~\ref{Ep-Lp-Eiso-z}, we show the $E_p$--$L_p$ (upper left) 
and the $E_p$--$E_{\rm iso}$ (upper right) correlations
of each class. The best-fit results and 1~$\sigma$ statistical 
errors are summarized in table~\ref{fit-redshift}. 
The 2~$\sigma$ confidence regions for three classes are also shown 
in the bottom left and right panels in figure~\ref{Ep-Lp-Eiso-z}, 
respectively. The meaning of each solid line is the same as one of 
figure~\ref{Ep-Lp-Eiso-Flux}. 

A difference between high- and low-redshift classes can be seen in 
the $E_p$--$L_p$ correlation at 2~$\sigma$ level, and a systematic 
redshift dependence is also seen in the power-law index.
Especially the discrepancy is remarkable toward higher value of 
$E_{p}$ and $L_{p}$. 
By contrast, in the $E_p$--$E_{\rm iso}$ correlation, 
the three classes are consistent at 1~$\sigma$ level through 
the entire correlations.

\section{Discussion and Summary} \label{sec:discussion}

In this paper, combining data from multiple detectors and recalculating
the peak flux for Konus events, we constructed a uniform database of
GRBs with known redshift and well-observed spectral quantities
to determine $E_p$, $L_p$ and $E_{\rm iso}$.
The 101~GRB samples enabled us to derive the $E_p$--$L_p$
and $E_p$--$E_{\rm iso}$ correlations with small statistical
errors compared with their dispersions. By dividing the samples
according to the observed flux/fluence and showing the correlation
of each group, it was shown that the correlations are intrinsic
to GRBs and not due to the truncation effect of the detector threshold.
Then we examined the flux/fluence dependence and redshift evolution
of the correlations, which is a crucial issue when we use GRBs
as cosmological tools. We found a fluence-dependence
in the $E_p$--$E_{\rm iso}$ correlation, and a redshift
dependence in the $E_p$--$L_p$ correlation. Both dependences
are about 2~$\sigma$ statistical level and relatively weak so
that they could still be used as cosmological tools. However,
they would contribute to the data dispersion in the correlations.


Let us give some comments on the work by \citet{butler}.
They suggested, using the {\it Swift} data,
the $E_p$--$E_{\rm iso}$ correlation is due to the detector
threshold. The $E^{\rm obs}_p$ value of many {\it Swift}
events can not be determined by the observational data because
of the narrow energy range of BAT instrument. So, using 
Bayesian statistics approach, they estimated the $E^{\rm obs}_p$ 
values for 218~events based on a parent distribution of spectral 
parameters suggested by BATSE data \citep{preece00}. 
\citet{butler} found a pseudo $E_p$--$E_{\rm iso}$ correlation
which is inconsistent with the original one by \citet{amati02}.
They claimed that this inconsistency is caused by the large
intrinsic scatter of the correlation.
In other words, the ``tight'' $E_p$--$E_{\rm iso}$ correlation
is an artifact result, because almost all dimmer samples 
close to the detector sensitivity would be outliers of 
the $E_p$--$E_{\rm iso}$ correlation by \citet{amati02}.
However we need caution to compare their work with others,
because their $E_p^{\rm obs}$ values are just estimated
by the Bayesian statistics approach, and not observed.
These simulated $E_p^{\rm obs}$ values highly depend
on the assumed parent distribution of spectral parameters.
Our approach to prove the existence of the correlation
is very different from this approach. We do not assume anything
about the statistical property of GRB samples.

The fluence dependence in the $E_p$--$E_{\rm iso}$ correlation
may be consistent with the conclusion by \citet{nava08},
although the statistical significance is only 2~$\sigma$ level.
This dependence might be interpreted as follows. We can observe only 
bright parts in the lightcurve for dim GRBs and tend to underestimate 
their total energies. In other words, the dispersion of 
the $E_p$--$E_{\rm iso}$ correlation might be influenced by 
the instrumental threshold effect because the dimmer events can be 
observed only for the more sensitive instruments.
In contrast, this effect would not be expected for the $E_p$--$L_p$
correlation because it involves the peak luminosity ($L_p$) which 
is determined only by the brightest part of each event.

We also found a weak redshift dependence in the $E_p$--$L_p$ 
correlation. Of course, one interprets that this dependence 
is the intrinsic property of GRBs. However, let us point out
a possibility of systematic overestimation of the peak luminosity
$L_p$ for high-redshift GRBs. In the $E_p$--$L_p$ correlation,
we usually use 1~second peak flux in the observer frame when we 
estimate $L_p$.
However this means that the time scale of peak luminosity $L_p$
in the GRB frame depends on the redshift because of the cosmological
time dilation. In other words, the time scale in the GRB frame
becomes shorter for higher-redshift GRBs. In general, we can expect
the peak luminosity increases as the time scale becomes shorter
as shown in figure~\ref{peak-flux}. For example, considering
a GRB at $z = 4$, the observed 1~second peak flux corresponds
to the 200~msec one at the GRB frame. According to
figure~\ref{peak-flux}, in this case, we systematically
overestimate the peak luminosity about 30--50~\% on average.

Thus, difference in the time interval in the GRB frame
would induce a systematic error in $L_p$. This effect would make
high-redshift GRBs look systematically brighter than low-redshift
GRBs and may result in the apparent redshift dependence suggested
in figure~\ref{Ep-Lp-Eiso-z}. Therefore, the redshift dependence
of the $E_p$--$L_p$ correlation might be understood as a selection
bias due to the inappropriate definition of $L_p$, rather than 
an intrinsic property of GRBs. If this argument is true, 
we may enable to reduce the systematic errors of $E_p$--$L_p$ 
correlation when we use the same time scale in each GRB frame.
Then we will make the $E_p$--$L_p$ correlation more accurate 
luminosity indicator. This will be presented elsewhere in near future.

\section*{Acknowledgments}

This work is supported in part by the Grant-in-Aid from the 
Ministry of Education, Culture, Sports, Science and Technology
(MEXT) of Japan, No.18684007 (DY), No.19540283, No.19047004(TN),
and No.21840028(KT), and by the Grant-in-Aid
for the global COE program {\it The Next Generation of Physics,
Spun from Universality and Emergence} at Kyoto University and
"Quest for Fundamental Principles in the Universe: from Particles
to the Solar System and the Cosmos" at Nagoya University
from MEXT of Japan. RT is supported by a Grant-in-Aid for
the Japan Society for the Promotion of Science (JSPS) Fellows
and is a research fellow of JSPS.

\clearpage
\begin{table}
\caption{The best fit results of the $E_p$--$L_p$ correlation and
the $E_p$--$E_{\rm iso}$ correlation in different flux (fluence) range.}
\begin{tabular}{lcccc}
\hline
Class  & $E_p$--$L_p$ & $\sigma_{sys, \log L_{p}}$ & $E_p$--$E_{\rm iso}$ & $\sigma_{sys, \log E_{\rm iso}}$ \\
\hline
dim    & $L_p = 10^{52.51\pm0.08} [E_{p,355}]^{1.82\pm0.15}$ & 0.23
       & $E_{\rm iso} = 10^{52.82\pm0.08} [E_{p,355}]^{1.63\pm0.15}$
       & 0.24 \\
middle & $L_p = 10^{52.40\pm0.06} [E_{p,355}]^{1.48\pm0.16}$ & 0.35
       & $E_{\rm iso} = 10^{52.96\pm0.05} [E_{p,355}]^{1.24\pm0.13}$
       & 0.29 \\
bright & $L_p = 10^{52.46\pm0.08} [E_{p,355}]^{1.53\pm0.17}$ & 0.35
       & $E_{\rm iso} = 10^{53.30\pm0.11} [E_{p,355}]^{1.29\pm0.25}$
       & 0.43 \\ 
\hline
\end{tabular}
\label{fit-flux}
\end{table}

\begin{table}
\caption{The definition of three different flux and fluence ranges.}
\begin{tabular}{lcc}
\hline
Class        & $F_{\rm p, bol}$ range         & $S_{\rm bol}$ range \\
             & $({\rm erg~cm^{-2}s^{-1}})$    & $({\rm erg~cm^{-2}})$\\
\hline
dim    & $F_{\rm p,bol} < 5 \times 10^{-7}$
       & $S_{\rm bol} < 5 \times 10^{-6}$\\
middle & $5 \times 10^{-7} \le F_{\rm p,bol} \le 4 \times 10^{-6}$
       & $5 \times 10^{-6} \le S_{\rm bol} \le 3 \times 10^{-5}$ \\
bright & $F_{\rm p,bol} > 4 \times 10^{-6}$
       & $S_{\rm bol} > 3 \times 10^{-5}$\\
\hline
\end{tabular}
\label{table-flux}
\end{table}

\begin{table}
\caption{The best fit results of the $E_p$--$L_p$ and
$E_p$--$E_{\rm iso}$ correlations for three redshift ranges.}
\begin{tabular}{lcccc}
\hline
Class     & $E_p$--$L_p$ & $\sigma_{sys}$ & $E_p$--$E_{\rm iso}$
          & $\sigma_{sys}$ \\
\hline
$z<1$     & $L_p = 10^{52.18\pm0.07} [E_{p,355}]^{1.33\pm0.13}$ & 0.32
          & $E_{\rm iso} = 10^{53.10\pm0.09} [E_{p,355}]^{1.74\pm0.15}$
          & 0.40 \\
$1<z<2.6$ & $L_p = 10^{52.47\pm0.06} [E_{p,355}]^{1.54\pm0.15}$ & 0.30
          & $E_{\rm iso} = 10^{52.97\pm0.08} [E_{p,355}]^{1.44\pm0.21}$
          & 0.43 \\
$z>2.6$   & $L_p = 10^{52.58\pm0.06} [E_{p,355}]^{1.78\pm0.17}$ & 0.23
          & $E_{\rm iso} = 10^{53.05\pm0.07} [E_{p,355}]^{1.70\pm0.20}$
          & 0.31 \\
\hline
\end{tabular}
\label{fit-redshift}
\end{table}

\clearpage
\appendix

{\tiny \tabcolsep=0.5mm
\begin{longtable}{lcccccccccc}
  \caption{Spectral parameters of 101~GRBs with known redshift, 
which is applied for appropriate $k$-correction.}
\label{tab:data1}
\hline              
    &          &          &         & $E^{obs}_p$ & $F_{p,bol}$$^{*}$        & $S_{\rm bol}$$^{*}$  & $L_p$$^{\dagger}$ & $E_{\rm iso}$$^{\dagger}$ & $T^{obs}_{90}/(1+z)$$^{\flat}$\\ 
GRB & redshift & $\alpha$ & $\beta$ & (keV)       & (erg~cm$^{-2}$~s$^{-1}$) & (erg~cm$^{-2}$)      & (erg~s$^{-1}$)    & (erg)               & (sec)\\ 
\endfirsthead
\hline
    &          &          &         & $E^{obs}_p$ & $F_{p,bol}$$^{*}$        & $S_{\rm bol}$$^{*}$  & $L_p$$^{\dagger}$ & $E_{\rm iso}$$^{\dagger}$ & $T^{obs}_{90}/(1+z)$$^{\flat}$\\ 
GRB & redshift & $\alpha$ & $\beta$ & (keV)       & (erg~cm$^{-2}$~s$^{-1}$) & (erg~cm$^{-2}$)      & (erg~s$^{-1}$)    & (erg)               & (sec)\\ 
\hline
\endhead
\hline
\multicolumn{11}{l}{$*$~Integrated between $1/(1+z)$--$10,000/(1+z)$~keV 
in the observer frame.}\\
\multicolumn{11}{l}{$\dagger$~Integrated between 1--10,000~keV 
in the GRB frame.}\\
\multicolumn{11}{l}{$\sharp$~Short GRB with 
$T_{90}^{obs}/(1+z) \le 1~{\rm sec}$ in the rest frame of GRB 
and/or $E_{p} > 100~{\rm keV}$ \citep{levesque2010}. Green points in Fig.1.}\\
\multicolumn{11}{l}{$\natural$~Merginal short GRB with 
$T_{90}^{obs}/(1+z) \le 2~{\rm sec}$ in the rest frame of GRB
but $E_{p} < 100~{\rm keV}$. Green points in Fig.1.}\\
\multicolumn{11}{l}{$\ddagger$~High-redshift GRB with the redshift of $z > 6$. 
Blue points in Fig.1.}\\
\multicolumn{11}{l}{$\flat$~$T_{90}^{obs}$ highly depends on 
the energy band width of each instruments.}\\
\endfoot
\hline
\multicolumn{11}{l}{$*$~Integrated between $1/(1+z)$--$10,000/(1+z)$~keV 
in the observer frame.}\\
\multicolumn{11}{l}{$\dagger$~Integrated between 1--10,000~keV 
in the GRB frame.}\\
\multicolumn{11}{l}{$\sharp$~Short GRB with 
$T_{90}^{obs}/(1+z) \le 1~{\rm sec}$ in the rest frame of GRB 
and/or $E_{p} > 100~{\rm keV}$ \citep{levesque2010}. Green points in Fig.1.}\\
\multicolumn{11}{l}{$\natural$~Merginal short GRB with 
$T_{90}^{obs}/(1+z) \le 2~{\rm sec}$ in the rest frame of GRB
but $E_{p} < 100~{\rm keV}$. Green points in Fig.1.}\\
\multicolumn{11}{l}{$\ddagger$~High-redshift GRB with the redshift of $z > 6$. 
Blue points in Fig.1.}\\
\multicolumn{11}{l}{$\flat$~$T_{90}^{obs}$ highly depends on 
the energy band width of each instruments.}\\
\endlastfoot
  \hline
970228	&	0.695	& $	-1.54	^{+	0.08	}_{	-0.08	}$ & $	-2.5	^{+	0.4	}_{
-0.4	}$ & $	115.0 	^{+	38.0 	}_{	-38.0 	}$ & $(	7.35	^{+	0.99 	}_{
-0.60 	})\times 10^{	-6	}$ & $(	2.19 	^{+	0.44 	}_{	-0.31 	})\times
10^{	-5	}$ & $(	1.54 	^{+	0.21 	}_{	-0.12 	})\times 10^{	52	}$ & $(
2.70 	^{+	0.55 	}_{	-0.39 	})\times 10^{	52	} $ & 47.2 \\
970508	&	0.835	& $	-1.03	^{+	1.51	}_{	-0.06	}$ & $	-2.2	^{+	0.1	}_{
-0.11	}$ & $	48.9 	^{+	20.6 	}_{	-16.2 	}$ & $(	5.03 	^{+	0.65 	}_{
-0.63 	})\times 10^{	-7	}$ & $(	5.51 	^{+	0.68 	}_{	-0.82 	})\times
10^{	-6	}$ & $(	1.64 	^{+	0.21 	}_{	-0.21 	})\times 10^{	51	}$ & $(
9.78 	^{+	1.20 	}_{	-1.46 	})\times 10^{	51	} $ & 12.6\\
970828	&	0.957	& $	-0.7	^{+	0.08	}_{	-0.08	}$ & $	-2.07	^{+	0.37	}_{
-0.37	}$ & $	298.0 	^{+	59.0 	}_{	-59.0 	}$ & $(	6.04 	^{+	0.34 	}_{
-0.34 	})\times 10^{	-6	}$ & $(	6.94 	^{+	0.29 	}_{	-0.29 	})\times
10^{	-4	}$ & $(	2.77 	^{+	0.16 	}_{	-0.16 	})\times 10^{	52	}$ & $(
1.63 	^{+	0.07 	}_{	-0.07 	})\times 10^{	54	} $ & 74.9\\
971214	&	3.42	& $	-0.36	^{+	0.14	}_{	-0.14	}$ & $	-2.1	^{+	0.52	}_{
-0.9	}$ & $	182.6 	^{+	11.0 	}_{	-14.3 	}$ & $(	9.22 	^{+	5.10 	}_{
-3.84 	})\times 10^{	-7	}$ & $(	1.32 	^{+	0.11 	}_{	-0.10 	})\times
10^{	-5	}$ & $(	9.49 	^{+	5.25 	}_{	-3.95 	})\times 10^{	52	}$ & $(
3.07 	^{+	0.25 	}_{	-0.23 	})\times 10^{	53	} $ & 7.1  \\
980613	&	1.096	& $	-1.43	^{+	0.24	}_{	-0.24	}$ & $	-2.7	^{+	0.6	}_{
-0.6	}$ & $	93.0 	^{+	43.0 	}_{	-43.0 	}$ & $(	3.04 	^{+	1.07 	}_{
-0.89 	})\times 10^{	-7	}$ & $(	1.90 	^{+	0.57 	}_{	-0.46 	})\times
10^{	-6	}$ & $(	1.93 	^{+	0.68 	}_{	-0.57 	})\times 10^{	51	}$ & $(
5.77 	^{+	1.72 	}_{	-1.41 	})\times 10^{	51	} $ & 9.5  \\
990123	&	1.6	& $	-0.18	^{+	0.08	}_{	-0.07	}$ & $	-2.33	^{+	0.08	}_{
-0.09	}$ & $	513.0 	^{+	19.2 	}_{	-21.9 	}$ & $(	1.70 	^{+	0.07 	}_{
-0.07 	})\times 10^{	-5	}$ & $(	5.72 	^{+	0.11 	}_{	-0.12 	})\times
10^{	-4	}$ & $(	2.75 	^{+	0.12 	}_{	-0.12 	})\times 10^{	53	}$ & $(
3.56 	^{+	0.07 	}_{	-0.07 	})\times 10^{	54	} $ & 24.4 \\
990506	&	1.3	& $	-0.9	^{+	0.19	}_{	-0.13	}$ & $	-2.08	^{+	0.08	}_{	-0.1
}$ & $	320.7 	^{+	30.1 	}_{	-38.2 	}$ & $(	1.23 	^{+	0.08 	}_{	-0.09
})\times 10^{	-5	}$ & $(	2.62 	^{+	0.08 	}_{	-0.09 	})\times 10^{	-4
}$ & $(	1.19 	^{+	0.08 	}_{	-0.09 	})\times 10^{	53	}$ & $(	1.11
^{+	0.03 	}_{	-0.04 	})\times 10^{	54	} $ & 56.5\\
990510	&	1.619	& $	-0.71	^{+	0.12	}_{	-0.12	}$ & $	-2.79	^{+	0.51	}_{
-6.21	}$ & $	205.5 	^{+	9.6 	}_{	-12.3 	}$ & $(	3.11 	^{+	0.64 	}_{
-0.86 	})\times 10^{	-6	}$ & $(	2.25 	^{+	0.17 	}_{	-0.17 	})\times
10^{	-5	}$ & $(	5.21 	^{+	1.08 	}_{	-1.44 	})\times 10^{	52	}$ & $(
1.44 	^{+	0.11 	}_{	-0.11 	})\times 10^{	53	} $ & 26.0\\
990705	&	0.843	& $	-1.05	^{+	0.21	}_{	-0.21	}$ & $	-2.2	^{+	0.1	}_{
-0.1	}$ & $	189.0 	^{+	15.0 	}_{	-15.0 	}$ & $(	6.61 	^{+	0.56 	}_{
-0.52 	})\times 10^{	-6	}$ & $(	1.34 	^{+	0.23 	}_{	-0.21 	})\times
10^{	-4	}$ & $(	2.21 	^{+	0.19 	}_{	-0.17 	})\times 10^{	52	}$ & $(
2.43 	^{+	0.41 	}_{	-0.38 	})\times 10^{	53	} $ & 22.8\\
990712	&	0.43	& $	-1.88	^{+	0.07	}_{	-0.07	}$ & $	-2.48	^{+	0.56	}_{
-0.56	}$ & $	65.0 	^{+	11.0 	}_{	-11.0 	}$ & $(	3.47 	^{+	0.68 	}_{
-0.51 	})\times 10^{	-6	}$ & $(	1.74 	^{+	0.28 	}_{	-0.20 	})\times
10^{	-5	}$ & $(	2.27 	^{+	0.45 	}_{	-0.33 	})\times 10^{	51	}$ & $(
7.97 	^{+	1.30 	}_{	-0.92 	})\times 10^{	51	} $ & 14.0\\
991208	&	0.71	& $	-1.1	^{+	0.4	}_{	-0.4	}$ & $	-2.2	^{+	0.4	}_{	-0.4
}$ & $	190.0 	^{+	20.0 	}_{	-20.0 	}$ & $(	2.14 	^{+	0.56 	}_{	-0.38
})\times 10^{	-5	}$ & 	--- & $(	4.68 	^{+	1.22 	}_{	-0.83
})\times 10^{	52	}$ & 	---	& 35.1	\\
991216	&	1.02	& $	-0.66	^{+	0.04	}_{	-0.04	}$ & $	-2.44	^{+	0.12	}_{
-0.17	}$ & $	536.5 	^{+	18.5 	}_{	-20.4 	}$ & $(	5.80 	^{+	0.27 	}_{
-0.35 	})\times 10^{	-5	}$ & $(	3.00 	^{+	0.08 	}_{	-0.11 	})\times
10^{	-4	}$ & $(	3.09 	^{+	0.14 	}_{	-0.19 	})\times 10^{	53	}$ & $(
7.92 	^{+	0.21 	}_{	-0.30 	})\times 10^{	53	} $ & 7.5\\
000131	&	4.5	& $	-0.91	^{+	0.2	}_{	-0.15	}$ & $	-2.02	^{+	0.18	}_{
-0.32	}$ & $	168.4 	^{+	17.7 	}_{	-15.1 	}$ & $(	2.17 	^{+	0.33 	}_{
-0.33 	})\times 10^{	-6	}$ & $(	1.85 	^{+	0.02 	}_{	-0.02 	})\times
10^{	-4	}$ & $(	4.26 	^{+	0.64 	}_{	-0.64 	})\times 10^{	53	}$ & $(
6.60 	^{+	0.08 	}_{	-0.08 	})\times 10^{	54	} $ & 20.0\\
000210	&	0.85	& $	-1.1	^{+	0.4	}_{	-0.4	}$ & $	-4.9	^{+	0.4	}_{	-0.4
}$ & $	408.0 	^{+	14.0 	}_{	-14.0 	}$ & $(	2.04 	^{+	0.83 	}_{	-0.51
})\times 10^{	-5	}$ & $(	2.88 	^{+	1.17 	}_{	-0.72 	})\times 10^{	-4
}$ & $(	6.98 	^{+	2.84 	}_{	-1.75 	})\times 10^{	52	}$ & $(	5.32
^{+	2.16 	}_{	-1.33 	})\times 10^{	53	} $ & 8.6\\
010921	&	0.45	& $	-1.55	^{+	0.08	}_{	-0.07	}$ & $	-2.25	^{+	fix	}_{	-fix
}$ & $	88.6 	^{+	21.7 	}_{	-13.8 	}$ & $(	1.78 	^{+	0.13 	}_{	-0.13
})\times 10^{	-6	}$ & $(	2.65 	^{+	0.20 	}_{	-0.20 	})\times 10^{	-5
}$ & $(	1.30 	^{+	0.09 	}_{	-0.09 	})\times 10^{	51	}$ & $(	1.33
^{+	0.10 	}_{	-0.10 	})\times 10^{	52	} $ & 8.3\\
020124	&	3.198	& $	-0.79	^{+	0.15	}_{	-0.14	}$ & $	-2.25	^{+	fix	}_{
-fix	}$ & $	86.9 	^{+	18.1 	}_{	-12.5 	}$ & $(	5.93 	^{+	0.86 	}_{
-0.86 	})\times 10^{	-7	}$ & $(	1.06 	^{+	0.15 	}_{	-0.13 	})\times
10^{	-5	}$ & $(	5.19 	^{+	0.75 	}_{	-0.75 	})\times 10^{	52	}$ & $(
2.21 	^{+	0.32 	}_{	-0.28 	})\times 10^{	53	} $ & 18.7\\ 
020127	&	1.9	& $	-1.03	^{+	0.14	}_{	-0.13	}$ & $	-2.25	^{+	fix	}_{
-fix	}$ & $	104.0 	^{+	47.0 	}_{	-24.1 	}$ & $(	5.80 	^{+	1.10 	}_{
-1.25 	})\times 10^{	-7	}$ & $(	3.92 	^{+	0.64 	}_{	-0.52 	})\times
10^{	-6	}$ & $(	1.43 	^{+	0.27 	}_{	-0.31 	})\times 10^{	52	}$ & $(
3.33 	^{+	0.54 	}_{	-0.45 	})\times 10^{	52	} $ & 2.8\\
020405	&	0.69	& $	0	^{+	0.25	}_{	-0.25	}$ & $	-1.87	^{+	0.23	}_{
-0.23	}$ & $	364.0 	^{+	73.0 	}_{	-73.0 	}$ & $(	2.50 	^{+	0.80 	}_{
-0.58 	})\times 10^{	-5	}$ & $(	7.51 	^{+	2.39 	}_{	-1.74 	})\times
10^{	-4	}$ & $(	5.15 	^{+	1.64 	}_{	-1.20 	})\times 10^{	52	}$ & $(
9.15 	^{+	2.91 	}_{	-2.12 	})\times 10^{	53	} $ & 23.7\\
020813	&	1.25	& $	-0.94	^{+	0.03	}_{	-0.03	}$ & $	-2.25	^{+	fix
}_{	-fix	}$ & $	142.0 	^{+	14.0 	}_{	-13.0 	}$ & $(	3.75 	^{+	0.20 	}_{
-0.20 	})\times 10^{	-6	}$ & $(	1.59 	^{+	0.02 	}_{	-0.02 	})\times
10^{	-4	}$ & $(	3.31 	^{+	0.18 	}_{	-0.18 	})\times 10^{	52	}$ & $(
6.23 	^{+	0.09 	}_{	-0.09 	})\times 10^{	53	} $ & 55.6\\
020819	&	0.41	& $	-0.9	^{+	0.17	}_{	-0.14	}$ & $	-1.99	^{+	0.18	}_{
-0.48	}$ & $	49.9 	^{+	17.9 	}_{	-12.2 	}$ & $(	1.37 	^{+	0.66 	}_{
-0.82 	})\times 10^{	-6	}$ & $(	1.61 	^{+	0.63 	}_{	-0.80 	})\times
10^{	-5	}$ & $(	8.01 	^{+	3.86 	}_{	-4.78 	})\times 10^{	50	}$ & $(
6.67 	^{+	2.61 	}_{	-3.31 	})\times 10^{	51	} $ & 14.2\\
020903	&	0.25	& $	-1	^{+	fix	}_{	-fix	}$ & $	-2.6	^{+	0.4	}_{	-0.6
}$ & $	2.6 	^{+	1.4 	}_{	-0.8 	}$ & $(	2.94 	^{+	2.14 	}_{	-1.35
})\times 10^{	-8	}$ & $(	1.81 	^{+	1.17 	}_{	-0.55 	})\times 10^{	-7
}$ & $(	5.42 	^{+	3.94 	}_{	-2.49 	})\times 10^{	48	}$ & $(	2.67
^{+	1.73 	}_{	-0.82 	})\times 10^{	49	} $ & 2.6\\
021004	&	2.335	& $	-1	^{+	0.2	}_{	-0.2	}$ & $	-2.25	^{+	fix	}_{
-fix	}$ & $	80.0 	^{+	53.0 	}_{	-23.0 	}$ & $(	2.21 	^{+	0.47 	}_{
-0.47 	})\times 10^{	-7	}$ & $(	3.32 	^{+	1.58 	}_{	-0.93 	})\times
10^{	-6	}$ & $(	9.05 	^{+	1.93 	}_{	-1.93 	})\times 10^{	51	}$ & $(
4.08 	^{+	1.94 	}_{	-1.14 	})\times 10^{	52	} $ & 30.0\\
021211	&	1.01	& $	-0.86	^{+	0.1	}_{	-0.09	}$ & $	-2.18	^{+	0.14	}_{
-0.25	}$ & $	45.6 	^{+	7.8 	}_{	-6.2 	}$ & $(	1.63 	^{+	0.40 	}_{
-0.45 	})\times 10^{	-6	}$ & $(	5.07 	^{+	0.96 	}_{	-1.08 	})\times
10^{	-6	}$ & $(	8.52 	^{+	2.07 	}_{	-2.37 	})\times 10^{	51	}$ & $(
1.32 	^{+	0.25 	}_{	-0.28 	})\times 10^{	52	} $ & 2.8\\
030115	&	2.5	& $	-1.28	^{+	0.14	}_{	-0.14	}$ & $	-2.2	^{+	0.4	}_{
-0.4	}$ & $	83.0 	^{+	53.0 	}_{	-22.0 	}$ & $(	3.17 	^{+	0.61 	}_{
-0.27 	})\times 10^{	-7	}$ & $(	3.04 	^{+	1.00 	}_{	-0.64 	})\times
10^{	-6	}$ & $(	1.53 	^{+	0.30 	}_{	-0.13 	})\times 10^{	52	}$ & $(
4.19 	^{+	1.38 	}_{	-0.88 	})\times 10^{	52	} $ & 5.7 \\
030226	&	1.98	& $	-0.89	^{+	0.17	}_{	-0.15	}$ & $	-2.25	^{+	fix	}_{	-fix
}$ & $	97.1 	^{+	27.0 	}_{	-17.1 	}$ & $(	2.51 	^{+	0.26 	}_{	-0.26
})\times 10^{	-7	}$ & $(	7.81 	^{+	1.25 	}_{	-1.10 	})\times 10^{	-6
}$ & $(	6.88 	^{+	0.70 	}_{	-0.70 	})\times 10^{	51	}$ & $(	7.18
^{+	1.15 	}_{	-1.01 	})\times 10^{	52	} $ & 33.6\\
030323	&	3.372	& $	-0.8	^{+	0.8	}_{	-0.8	}$ & $	-2.25	^{+	fix	}_{
-fix	}$ & $	44.0 	^{+	90.0 	}_{	-26.0 	}$ & $(	1.15 	^{+	0.61 	}_{
-0.52 	})\times 10^{	-7	}$ & $(	6.58 	^{+	3.31 	}_{	-2.34 	})\times
10^{	-7	}$ & $(	1.14 	^{+	0.61 	}_{	-0.51 	})\times 10^{	52	}$ & $(
1.50 	^{+	0.75 	}_{	-0.53 	})\times 10^{	52	} $ & 5.9\\
030328	&	1.52	& $	-1.14	^{+	0.03	}_{	-0.03	}$ & $	-2.09	^{+	0.19
}_{	-0.4	}$ & $	126.3 	^{+	13.9 	}_{	-13.1 	}$ & $(	1.56 	^{+	0.40 	}_{
-0.47 	})\times 10^{	-6	}$ & $(	6.12 	^{+	1.29 	}_{	-1.59 	})\times
10^{	-5	}$ & $(	2.23 	^{+	0.56 	}_{	-0.67 	})\times 10^{	52	}$ & $(
3.47 	^{+	0.73 	}_{	-0.90 	})\times 10^{	53	} $ & 39.7 \\
030329	&	0.168	& $	-1.26	^{+	0.01	}_{	-0.02	}$ & $	-2.28	^{+
0.05	}_{	-0.06	}$ & $	67.9 	^{+	2.3 	}_{	-2.2 	}$ & $(	1.99 	^{+
0.16 	}_{	-0.16 	})\times 10^{	-5	}$ & $(	2.34 	^{+	0.09 	}_{	-0.10
})\times 10^{	-4	}$ & $(	1.51 	^{+	0.12 	}_{	-0.12 	})\times 10^{	51
}$ & $(	1.52 	^{+	0.06 	}_{	-0.06 	})\times 10^{	52	} $ & 42.8 \\
030429	&	2.65	& $	-1.12	^{+	0.25	}_{	-0.22	}$ & $	-2.25	^{+	fix	}_{	-fix
}$ & $	35.0 	^{+	11.8 	}_{	-7.9 	}$ & $(	1.35 	^{+	0.34 	}_{	-0.39
})\times 10^{	-7	}$ & $(	1.10 	^{+	0.19 	}_{	-0.17 	})\times 10^{	-6
}$ & $(	7.52 	^{+	1.88 	}_{	-2.15 	})\times 10^{	51	}$ & $(	1.68
^{+	0.29 	}_{	-0.26 	})\times 10^{	52	} $ & 3.8 \\
030528	&	0.782	& $	-1.33	^{+	0.15	}_{	-0.12	}$ & $	-2.65	^{+	0.29	}_{
-0.98	}$ & $	31.8 	^{+	4.7 	}_{	-5.0 	}$ & $(	4.37 	^{+	1.03 	}_{
-1.09 	})\times 10^{	-7	}$ & $(	1.38 	^{+	0.22 	}_{	-0.21 	})\times
10^{	-5	}$ & $(	1.22 	^{+	0.29 	}_{	-0.30 	})\times 10^{	51	}$ & $(
2.16 	^{+	0.34 	}_{	-0.32 	})\times 10^{	52	} $ & 33.7 \\
040924$^{\sharp}$	&	0.859	& $	-1	^{+	fix	}_{	-fix	}$ & $	-2.25	^{+	fix	}_{	-fix
}$ & $	67.0 	^{+	6.0 	}_{	-6.0 	}$ & $(	2.45 	^{+	0.26 	}_{	-0.26
})\times 10^{	-6	}$ & $(	4.59 	^{+	0.20 	}_{	-0.20 	})\times 10^{	-6
}$ & $(	8.58 	^{+	0.90 	}_{	-0.90 	})\times 10^{	51	}$ & $(	8.65
^{+	0.38 	}_{	-0.38 	})\times 10^{	51	} $ & 0.8 \\
041006	&	0.716	& $	-1.37	^{+	fix	}_{	-fix	}$ & $	-2.25	^{+	fix	}_{
-fix	}$ & $	63.0 	^{+	13.0 	}_{	-13.0 	}$ & $(	2.39 	^{+	0.10 	}_{
-0.10 	})\times 10^{	-6	}$ & 	---	& $(	5.31 	^{+	0.22 	}_{	-0.22
})\times 10^{	51	}$ & 	---  & 14.3 \\
050126	&	1.29	& $	-1.1	^{+	0.1	}_{	-0.1	}$ & $	-2.25	^{+	fix	}_{	-fix
}$ & $	47.0 	^{+	23.0 	}_{	-8.0 	}$ & $(	1.02 	^{+	0.25 	}_{	-0.25
})\times 10^{	-7	}$ & $(	1.90 	^{+	0.18 	}_{	-0.18 	})\times 10^{	-6
}$ & $(	9.69 	^{+	2.33 	}_{	-2.33 	})\times 10^{	50	}$ & $(	7.88
^{+	0.76 	}_{	-0.76 	})\times 10^{	51	} $ & 11.4\\
050315	&	1.949	& $	-1	^{+	fix	}_{	-fix	}$ & $	-2.04	^{+	0.16	}_{
-0.24	}$ & $	40.3 	^{+	8.5 	}_{	-11.4 	}$ & $(	3.32 	^{+	1.26 	}_{
-1.18 	})\times 10^{	-7	}$ & $(	8.61 	^{+	2.23 	}_{	-2.24 	})\times
10^{	-6	}$ & $(	8.75 	^{+	3.32 	}_{	-3.11 	})\times 10^{	51	}$ & $(
7.69 	^{+	1.99 	}_{	-2.00 	})\times 10^{	52	} $ & 32.6 \\
050318	&	1.44	& $	-1	^{+	fix	}_{	-fix	}$ & $	-2.1	^{+	0.11	}_{
-0.11	}$ & $	47.1 	^{+	10.2 	}_{	-10.2 	}$ & $(	5.27 	^{+	1.17 	}_{
-0.96 	})\times 10^{	-7	}$ & $(	2.76 	^{+	0.58 	}_{	-0.49 	})\times
10^{	-6	}$ & $(	6.57 	^{+	1.46 	}_{	-1.20 	})\times 10^{	51	}$ & $(
1.41 	^{+	0.30 	}_{	-0.25 	})\times 10^{	52	} $ & 13.1\\
050319	&	3.24	& $	-1	^{+	fix	}_{	-fix	}$ & $	-2.35	^{+	0.35	}_{	-0.35
}$ & $	70.0 	^{+	70.0 	}_{	-35.0 	}$ & $(	2.13 	^{+	0.94 	}_{	-0.62
})\times 10^{	-7	}$ & $(	2.66 	^{+	1.05 	}_{	-0.70 	})\times 10^{	-6
}$ & $(	1.92 	^{+	0.85 	}_{	-0.56 	})\times 10^{	52	}$ & $(	5.67
^{+	2.24 	}_{	-1.49 	})\times 10^{	52	} $ & 2.4 \\
050401	&	2.9	& $	-0.9	^{+	0.3	}_{	-0.3	}$ & $	-2.55	^{+	0.22	}_{
-0.44	}$ & $	117.5 	^{+	18.0 	}_{	-18.0 	}$ & $(	1.73 	^{+	0.33 	}_{
-0.37 	})\times 10^{	-6	}$ & $(	1.69 	^{+	0.20 	}_{	-0.25 	})\times
10^{	-5	}$ & $(	1.20 	^{+	0.23 	}_{	-0.26 	})\times 10^{	53	}$ & $(
3.01 	^{+	0.36 	}_{	-0.45 	})\times 10^{	53	} $ & 8.5 \\
050416A$^{\natural}$	&	0.6535	& $	-1	^{+	fix	}_{	-fix	}$ & $	-5.697	^{+	3.3	}_{
-4.2	}$ & $	16.4 	^{+	2.8 	}_{	-2.8 	}$ & $(	4.88 	^{+	1.05 	}_{	-0.84
})\times 10^{	-7	}$ & $(	8.82 	^{+	2.51 	}_{	-1.95 	})\times 10^{	-7
}$ & $(	8.68 	^{+	1.87 	}_{	-1.49 	})\times 10^{	50	}$ & $(	9.49
^{+	2.70 	}_{	-2.10 	})\times 10^{	50	} $ & 1.5\\
050502A	&	3.793	& $	-1.1	^{+	0.4	}_{	-0.4	}$ & $	-2.25	^{+	fix	}_{
-fix	}$ & $	93.0 	^{+	55.0 	}_{	-35.0 	}$ & $(	4.20 	^{+	1.05 	}_{
-1.05 	})\times 10^{	-7	}$ & 	---	& $(	5.49 	^{+	1.37 	}_{	-1.37
})\times 10^{	52	}$ &	---   & 4.2\\
050505  & 4.27  & $-0.95^{+0.31}_{-0.31} $ & $-2.25^{+fix}_{fix} $ & $125^{+46.5}_{-46.5} $ & $(2.51^{+0.34}_{-0.34}) \times 10^{-7} $ & $(6.52^{+0.64}_{-0.64}) \times 10^{-6} $ & $4.39^{+0.59}_{-0.65}) \times 10^{52} $ & $2.16^{+0.21}_{-0.21}) \times 10^{53} $ & 11.4\\
050525	&	0.606	& $	-1.01	^{+	0.11	}_{	-0.11	}$ & $	-3.26	^{+	0.23
}_{	-0.41	}$ & $	81.2 	^{+	2.3 	}_{	-2.3 	}$ & $(	4.72 	^{+	0.27 	}_{
-0.29 	})\times 10^{	-6	}$ & $(	2.43 	^{+	0.12 	}_{	-0.13 	})\times
10^{	-5	}$ & $(	7.05 	^{+	0.40 	}_{	-0.44 	})\times 10^{	51	}$ & $(
2.26 	^{+	0.11 	}_{	-0.12 	})\times 10^{	52	} $ & 5.5 \\
050603	&	2.821	& $	-1.03	^{+	0.11	}_{	-0.11	}$ & $	-2.03	^{+	0.17
}_{	-0.29	}$ & $	343.7 	^{+	87.0 	}_{	-87.0 	}$ & $(	8.00 	^{+	1.22
}_{	-1.36 	})\times 10^{	-6	}$ & $(	2.73 	^{+	0.38 	}_{	-0.43
})\times 10^{	-5	}$ & $(	5.18 	^{+	0.79 	}_{	-0.88 	})\times 10^{	53
}$ & $(	4.63 	^{+	0.64 	}_{	-0.72 	})\times 10^{	53	} $ & 2.6 \\
050814  & 5.3  & $-0.58^{+0.56}_{-0.56} $ & $-2.25^{+fix}_{fix} $ & $54^{+7.5}_{-7.5} $ & $(1.24^{+0.37}_{-0.38}) \times 10^{-7} $ & $(3.19^{+0.53}_{-0.53}) \times 10^{-6} $ & $3.60^{+1.08}_{-1.11}) \times 10^{52} $ & $1.47^{+0.24}_{-0.24}) \times 10^{53} $ & 4.0\\
050820A	&	2.612	& $	-1.25	^{+	0.15	}_{	-0.15	}$ & $	-2.25	^{+	fix
}_{	-fix	}$ & $	246.0 	^{+	127.0 	}_{	-66.0 	}$ & $(	5.99 	^{+	0.56 	}_{
-0.83 	})\times 10^{	-7	}$ & $(	1.07 	^{+	0.08 	}_{	-0.10 	})\times
10^{	-5	}$ & $(	3.22 	^{+	0.30 	}_{	-0.45 	})\times 10^{	52	}$ & $(
1.59 	^{+	0.11 	}_{	-0.15 	})\times 10^{	53	} $ & 7.2 \\
050908	&	3.344	& $	-1	^{+	fix	}_{	-fix	}$ & $	-2.25	^{+	fix	}_{	-fix
}$ & $	41.0 	^{+	9.0 	}_{	-5.0 	}$ & $(	9.28 	^{+	0.14 	}_{	-1.86
})\times 10^{	-8	}$ & $(	1.03 	^{+	0.13 	}_{	-0.12 	})\times 10^{	-6
}$ & $(	9.05 	^{+	0.13 	}_{	-1.81 	})\times 10^{	51	}$ & $(	2.31
^{+	0.29 	}_{	-0.26 	})\times 10^{	52	} $ & 4.6 \\
050922C$^{\sharp}$	&	2.198	& $	-0.95	^{+	0.11	}_{	-0.11	}$ & $	-2.25	^{+	fix
}_{	-fix	}$ & $	196.8 	^{+	64.0 	}_{	-37.0 	}$ & $(	1.87 	^{+	0.08 	}_{
-0.15 	})\times 10^{	-6	}$ & $(	4.93 	^{+	0.22 	}_{	-0.42 	})\times
10^{	-6	}$ & $(	6.60 	^{+	0.29 	}_{	-0.54 	})\times 10^{	52	}$ & $(
5.44 	^{+	0.24 	}_{	-0.46 	})\times 10^{	52	} $ & 1.4 \\
051016B	&	0.9364	& $	-1	^{+	fix	}_{	-fix	}$ & $	-2.56	^{+	0.5	}_{
-6.33	}$ & $	27.5 	^{+	13.2 	}_{	-12.3 	}$ & $(	1.52 	^{+	1.11 	}_{
-0.87 	})\times 10^{	-7	}$ & $(	3.38 	^{+	1.46 	}_{	-1.16 	})\times
10^{	-7	}$ & $(	6.54 	^{+	4.77 	}_{	-3.75 	})\times 10^{	50	}$ & $(
7.51 	^{+	3.24 	}_{	-2.58 	})\times 10^{	50	} $ & 2.1 \\
051022	&	0.8	& $	-1.22	^{+	0.02	}_{	-0.02	}$ & $	-2.25	^{+	fix
}_{	-fix	}$ & $	306.0 	^{+	31.0 	}_{	-26.0 	}$ & $(	1.06 	^{+	0.14 	}_{
-0.14 	})\times 10^{	-5	}$ & $(	3.00 	^{+	0.04 	}_{	-0.04 	})\times
10^{	-4	}$ & $(	3.12 	^{+	0.41 	}_{	-0.41 	})\times 10^{	52	}$ & $(
4.90 	^{+	0.06 	}_{	-0.06 	})\times 10^{	53	} $ & 111.1 \\
051109A	&	2.346	& $	-1.38	^{+	0.33	}_{	-0.33	}$ & $	-2.25	^{+	fix
}_{	-fix	}$ & $	139.5 	^{+	116.0 	}_{	-45.0 	}$ & $(	7.47 	^{+	1.31 	}_{
-1.67 	})\times 10^{	-7	}$ & $(	5.77 	^{+	0.71 	}_{	-0.71 	})\times
10^{	-6	}$ & $(	3.09 	^{+	0.54 	}_{	-0.69 	})\times 10^{	52	}$ & $(
7.14 	^{+	0.88 	}_{	-0.88 	})\times 10^{	52	} $ & 10.8 \\
060115	&	3.53	& $	-1	^{+	0.5	}_{	-0.5	}$ & $	-2.25	^{+	fix	}_{	-fix	}$
& $	62.0 	^{+	31.0 	}_{	-10.0 	}$ & $(	1.25 	^{+	0.17 	}_{	-0.17
})\times 10^{	-7	}$ & $(	3.63 	^{+	0.32 	}_{	-0.32 	})\times 10^{	-6
}$ & $(	1.39 	^{+	0.19 	}_{	-0.19 	})\times 10^{	52	}$ & $(	8.90
^{+	0.78 	}_{	-0.78 	})\times 10^{	52	} $ & 31.3 \\
060124	&	2.296	& $	-1.29	^{+	0.14	}_{	-0.11	}$ & $	-2.25	^{+
0.27	}_{	-0.88	}$ & $	238.0 	^{+	126.0 	}_{	-85.0 	}$ & $(	2.61
^{+	0.79 	}_{	-0.75 	})\times 10^{	-6	}$ & $(	3.37 	^{+	0.08 	}_{
-0.11 	})\times 10^{	-5	}$ & $(	1.02 	^{+	0.31 	}_{	-0.29 	})\times
10^{	53	}$ & $(	4.01 	^{+	0.09 	}_{	-0.13 	})\times 10^{	53	} $ & 215.4 \\
060206$^{\natural}$	&	4.048	& $	-1.06	^{+	0.34	}_{	-0.34	}$ & $	-2.25	^{+	fix	}_{
-fix	}$ & $	75.4 	^{+	19.5 	}_{	-19.5 	}$ & $(	4.20 	^{+	0.26 	}_{
-0.26 	})\times 10^{	-7	}$ & $(	1.79 	^{+	0.09 	}_{	-0.09 	})\times
10^{	-6	}$ & $(	6.50 	^{+	0.40 	}_{	-0.40 	})\times 10^{	52	}$ & $(
5.49 	^{+	0.28 	}_{	-0.28 	})\times 10^{	52	} $ & 1.4 \\
060210	&	3.91	& $	-1.18	^{+	0.31	}_{	-0.31	}$ & $	-2.25	^{+	fix
}_{	-fix	}$ & $	149.0 	^{+	400.0 	}_{	-35.0 	}$ & $(	5.24 	^{+	1.40 	}_{
-0.93 	})\times 10^{	-7	}$ & $(	1.92 	^{+	0.64 	}_{	-0.10 	})\times
10^{	-5	}$ & $(	7.40 	^{+	1.98 	}_{	-1.32 	})\times 10^{	52	}$ & $(
5.52 	^{+	1.84 	}_{	-0.30 	})\times 10^{	53	} $ & 51.9 \\
060223A$^{\natural}$	&	4.41	& $	-1.18	^{+	0.31	}_{	-0.31	}$ & $	-2.25	^{+	fix
}_{	-fix	}$ & $	71.0 	^{+	100.0 	}_{	-10.0 	}$ & $(	2.00 	^{+	0.27 	}_{
-0.31 	})\times 10^{	-7	}$ & $(	1.46 	^{+	0.57 	}_{	-0.10 	})\times
10^{	-6	}$ & $(	3.74 	^{+	0.50 	}_{	-0.58 	})\times 10^{	52	}$ & $(
5.05 	^{+	1.97 	}_{	-0.36 	})\times 10^{	52	} $ & 2.0 \\
060510B	&	4.9	& $	-1.47	^{+	0.18	}_{	-0.18	}$ & $	-2.25	^{+	fix
}_{	-fix	}$ & $	95.0 	^{+	60.0 	}_{	-30.0 	}$ & $(	9.30 	^{+	1.80 	}_{
-1.80 	})\times 10^{	-8	}$ & $(	9.77 	^{+	0.58 	}_{	-0.52 	})\times
10^{	-6	}$ & $(	2.26 	^{+	0.44 	}_{	-0.44 	})\times 10^{	52	}$ & $(
4.02 	^{+	0.24 	}_{	-0.22 	})\times 10^{	53	} $ & 46.8 \\
060522  & 5.11  & $-0.7^{+0.44}_{-0.44} $ & $-2.25^{+fix}_{fix} $ & $70^{+13}_{-13} $ & $(8.69^{+2.90}_{-3.04}) \times 10^{-8} $ & $(2.21^{+0.20}_{-0.20}) \times 10^{-6} $ & $2.33^{+0.78}_{-0.81}) \times 10^{52} $ & $9.69^{+0.88}_{-0.88}) \times 10^{52} $ & 11.3\\
060526	&	3.221	& $	-1.1	^{+	fix	}_{	-fix	}$ & $	-2.25	^{+	fix	}_{	-fix
}$ & $	25.0 	^{+	5.0 	}_{	-5.0 	}$ & $(	2.24 	^{+	0.29 	}_{	-0.26
})\times 10^{	-7	}$ & $(	2.93 	^{+	0.52 	}_{	-0.47 	})\times 10^{	-6
}$ & $(	2.00 	^{+	0.26 	}_{	-0.24 	})\times 10^{	52	}$ & $(	6.18
^{+	1.10 	}_{	-1.00 	})\times 10^{	52	} $ & 3.3 \\
060604	&	2.68	& $	-1.34	^{+	0.3	}_{	-0.3	}$ & $	-2.25	^{+	fix	}_{	-fix
}$ & $	40.0 	^{+	5.0 	}_{	-5.0 	}$ & $(	8.66 	^{+	1.44 	}_{	-1.44
})\times 10^{	-8	}$ & $(	9.51 	^{+	2.51 	}_{	-2.51 	})\times 10^{	-7
}$ & $(	4.96 	^{+	0.82 	}_{	-0.82 	})\times 10^{	51	}$ & $(	1.48
^{+	0.39 	}_{	-0.39 	})\times 10^{	52	} $ & 2.7 \\
060707	&	3.425	& $	-0.6	^{+	0.7	}_{	-0.6	}$ & $	-2.25	^{+	fix	}_{	-fix
}$ & $	63.0 	^{+	21.0 	}_{	-10.0 	}$ & $(	1.46 	^{+	0.33 	}_{	-0.35
})\times 10^{	-7	}$ & $(	3.28 	^{+	0.31 	}_{	-0.31 	})\times 10^{	-6
}$ & $(	1.51 	^{+	0.34 	}_{	-0.36 	})\times 10^{	52	}$ & $(	7.66
^{+	0.72 	}_{	-0.72 	})\times 10^{	52	} $ & 15.4 \\
060714  & 2.711  & $-1.77^{+0.24}_{-0.24} $ & $-2.25^{+fix}_{fix} $ & $63^{+29}_{-29} $ & $(2.55^{+0.27}_{-0.30}) \times 10^{-7} $ & $(8.71^{+1.43}_{-0.58}) \times 10^{-6} $ & $1.50^{+0.16}_{-0.18}) \times 10^{52} $ & $1.38^{+0.23}_{-0.09}) \times 10^{53} $ & \\
060814  & 0.703  & $-1.43^{+0.15}_{-0.16} $ & $-2.25^{+fix}_{fix} $ & $257^{+122}_{-58} $ & $(2.91^{+0.40}_{-0.75}) \times 10^{-6} $ & $(4.18^{+0.19}_{-1.23}) \times 10^{-5} $ & $6.18^{+0.84}_{-1.60}) \times 10^{51} $ & $5.21^{+0.23}_{-1.53}) \times 10^{52} $ & 85.7\\
060908	&	2.43	& $	-1	^{+	0.3	}_{	-0.3	}$ & $	-2.25	^{+	fix	}_{	-fix
}$ & $	151.0 	^{+	184.0 	}_{	-41.0 	}$ & $(	6.48 	^{+	0.72 	}_{	-3.71
})\times 10^{	-7	}$ & $(	7.43 	^{+	1.32 	}_{	-2.30 	})\times 10^{	-6
}$ & $(	2.92 	^{+	0.32 	}_{	-1.67 	})\times 10^{	52	}$ & $(	9.77
^{+	1.73 	}_{	-3.02 	})\times 10^{	52	} $ & 5.6 \\
060927	&	5.6	& $	-0.9	^{+	0.4	}_{	-0.4	}$ & $	-2.25	^{+	fix	}_{	-fix	}$
& $	72.0 	^{+	25.0 	}_{	-11.0 	}$ & $(	3.89 	^{+	0.25 	}_{	-0.25
})\times 10^{	-7	}$ & $(	2.30 	^{+	0.14 	}_{	-0.14 	})\times 10^{	-6
}$ & $(	1.28 	^{+	0.08 	}_{	-0.08 	})\times 10^{	53	}$ & $(	1.15
^{+	0.07 	}_{	-0.07 	})\times 10^{	53	} $ & 3.4 \\
061007	&	1.261	& $	-0.7	^{+	0.04	}_{	-0.04	}$ & $	-2.61	^{+
0.15	}_{	-0.21	}$ & $	399.0 	^{+	19.0 	}_{	-18.0 	}$ & $(	1.59 	^{+
0.25 	}_{	-0.20 	})\times 10^{	-5	}$ & $(	2.39 	^{+	0.16 	}_{	-0.12
})\times 10^{	-4	}$ & $(	1.43 	^{+	0.23 	}_{	-0.18 	})\times 10^{	53
}$ & $(	9.54 	^{+	0.65 	}_{	-0.46 	})\times 10^{	53	} $ & 33.2 \\
070125	&	1.547	& $	-1.1	^{+	0.1	}_{	-0.09	}$ & $	-2.08	^{+	0.1
}_{	-0.15	}$ & $	367.0 	^{+	65.0 	}_{	-51.0 	}$ & $(	1.23 	^{+	0.20
}_{	-0.19 	})\times 10^{	-5	}$ & $(	1.52 	^{+	0.16 	}_{	-0.13
})\times 10^{	-4	}$ & $(	1.84 	^{+	0.29 	}_{	-0.28 	})\times 10^{	53
}$ & $(	8.92 	^{+	0.92 	}_{	-0.77 	})\times 10^{	53	} $ & 27.5 \\
070508	&	0.82	& $	-0.81	^{+	0.07	}_{	-0.07	}$ & $	-2.25	^{+
fix	}_{	-fix	}$ & $	188.0 	^{+	8.0 	}_{	-8.0 	}$ & $(	8.11 	^{+	1.01
}_{	-1.08 	})\times 10^{	-6	}$ & $(	5.58 	^{+	0.10 	}_{	-0.32
})\times 10^{	-5	}$ & $(	2.54 	^{+	0.32 	}_{	-0.34 	})\times 10^{	52
}$ & $(	9.62 	^{+	0.17 	}_{	-0.56 	})\times 10^{	52	} $ & 11.5 \\
070521	&	0.553	& $	-0.93	^{+	0.12	}_{	-0.12	}$ & $	-2.25	^{+
fix	}_{	-fix	}$ & $	222.0 	^{+	27.0 	}_{	-21.0 	}$ & $(	2.69 	^{+
0.51 	}_{	-0.70 	})\times 10^{	-6	}$ & $(	2.64 	^{+	0.09 	}_{	-0.45
})\times 10^{	-5	}$ & $(	3.21 	^{+	0.61 	}_{	-0.83 	})\times 10^{	51
}$ & $(	2.03 	^{+	0.07 	}_{	-0.35 	})\times 10^{	52	} $ & 24.4 \\
071003	&	1.60435	& $	-0.97	^{+	0.07	}_{	-0.07	}$ & $	-2.25
^{+	fix	}_{	-fix	}$ & $	799.0 	^{+	124.0 	}_{	-100.0 	}$ & $(	7.60
^{+	1.19 	}_{	-1.37 	})\times 10^{	-6	}$ & $(	5.37 	^{+	0.30 	}_{
-0.68 	})\times 10^{	-5	}$ & $(	1.24 	^{+	0.19 	}_{	-0.22 	})\times
10^{	53	}$ & $(	3.38 	^{+	0.19 	}_{	-0.43 	})\times 10^{	53	} $ & 11.5 \\
071010B	&	0.947	& $	-1.53	^{+	0.22	}_{	-0.22	}$ & $	-2.25	^{+
fix	}_{	-fix	}$ & $	52.0 	^{+	6.4 	}_{	-6.4 	}$ & $(	1.25 	^{+	0.05 	}_{
-0.06 	})\times 10^{	-6	}$ & $(	1.14 	^{+	0.09 	}_{	-0.06 	})\times
10^{	-5	}$ & $(	5.55 	^{+	0.22 	}_{	-0.26 	})\times 10^{	51	}$ & $(
2.60 	^{+	0.19 	}_{	-0.14 	})\times 10^{	52	} $ & 18.3 \\
071020$^{\sharp}$	&	2.145	& $	-0.65	^{+	0.27	}_{	-0.32	}$ & $	-2.25
^{+	fix	}_{	-fix	}$ & $	322.0 	^{+	80.0 	}_{	-53.0 	}$ & $(	4.13
^{+	0.77 	}_{	-2.31 	})\times 10^{	-6	}$ & $(	8.65 	^{+	0.44 	}_{
-5.34 	})\times 10^{	-6	}$ & $(	1.38 	^{+	0.26 	}_{	-0.77 	})\times
10^{	53	}$ & $(	9.17 	^{+	0.46 	}_{	-5.66 	})\times 10^{	52	} $ & 1.1 \\
071117	&	1.331	& $	-1.53	^{+	0.15	}_{	-0.16	}$ & $	-2.25
^{+	fix	}_{	-fix	}$ & $	278.0 	^{+	236.0 	}_{	-79.0 	}$ & $(	6.61
^{+	1.12 	}_{	-2.92 	})\times 10^{	-6	}$ & $(	9.11 	^{+	0.47 	}_{
-3.74 	})\times 10^{	-6	}$ & $(	6.79 	^{+	1.15 	}_{	-3.00 	})\times
10^{	52	}$ & $(	4.02 	^{+	0.21 	}_{	-1.65 	})\times 10^{	52	} $ & 2.1 \\
080319B	&	0.937	& $	-0.822	^{+	0.014	}_{	-0.012	}$ & $
-3.87	^{+	0.44	}_{	-1.09	}$ & $	651.0 	^{+	13.0 	}_{	-14.0 	}$ & $(
1.70 	^{+	0.16 	}_{	-0.16 	})\times 10^{	-5	}$ & $(	5.82 	^{+	0.14 	}_{
-0.13 	})\times 10^{	-4	}$ & $(	7.39 	^{+	0.71 	}_{	-0.71 	})\times
10^{	52	}$ & $(	1.31 	^{+	0.03 	}_{	-0.03 	})\times 10^{	54	} $ & 31.0 \\
080411	&	1.03	& $	-1.51	^{+	0.04	}_{	-0.05	}$ & $	-2.25	^{+
fix	}_{	-fix	}$ & $	259.0 	^{+	35.0 	}_{	-27.0 	}$ & $(	1.43 	^{+
0.18 	}_{	-0.18 	})\times 10^{	-5	}$ & $(	8.56 	^{+	0.42 	}_{	-0.40
})\times 10^{	-5	}$ & $(	7.85 	^{+	0.98 	}_{	-0.98 	})\times 10^{	52
}$ & $(	2.31 	^{+	0.11 	}_{	-0.11 	})\times 10^{	53	} $ & 34.5 \\
080413	&	2.433	& $	-1.2	^{+	0.1	}_{	-0.1	}$ & $	-2.25	^{+
fix	}_{	-fix	}$ & $	170.0 	^{+	80.0 	}_{	-40.0 	}$ & $(	9.88 	^{+
3.77 	}_{	-2.72 	})\times 10^{	-7	}$ & $(	5.02 	^{+	0.53 	}_{	-1.58
})\times 10^{	-6	}$ & $(	4.47 	^{+	1.71 	}_{	-1.23 	})\times 10^{	52
}$ & $(	6.61 	^{+	0.70 	}_{	-2.08 	})\times 10^{	52	} $ & 13.4 \\
080413B	&	1.1	& $	-1.26	^{+	0.27	}_{	-0.27	}$ & $	-2.25	^{+	fix
}_{	-fix	}$ & $	73.3 	^{+	15.8 	}_{	-15.8 	}$ & $(	3.05 	^{+	0.13 	}_{
-0.13 	})\times 10^{	-6	}$ & $(	7.70 	^{+	0.24 	}_{	-0.26 	})\times
10^{	-6	}$ & $(	1.96 	^{+	0.08 	}_{	-0.08 	})\times 10^{	52	}$ & $(
2.35 	^{+	0.07 	}_{	-0.08 	})\times 10^{	52	} $ & 3.8 \\
080603B	&	2.69	& $	-1.21	^{+	0.3	}_{	-0.3	}$ & $	-2.25	^{+	fix	}_{
-fix	}$ & $	71.0 	^{+	16.0 	}_{	-16.0 	}$ & $(	5.39 	^{+	0.31 	}_{
-0.31 	})\times 10^{	-7	}$ & $(	5.45 	^{+	0.23 	}_{	-0.23 	})\times
10^{	-6	}$ & $(	3.11 	^{+	0.18 	}_{	-0.18 	})\times 10^{	52	}$ & $(
8.53 	^{+	0.36 	}_{	-0.36 	})\times 10^{	52	} $ & 16.3 \\
080605	&	1.6398	& $	-1.03	^{+	0.07	}_{	-0.07	}$ & $	-2.25
^{+	fix	}_{	-fix	}$ & $	252.0 	^{+	20.0 	}_{	-17.0 	}$ & $(	1.11
^{+	0.23 	}_{	-0.23 	})\times 10^{	-5	}$ & $(	3.56 	^{+	0.15 	}_{
-0.14 	})\times 10^{	-5	}$ & $(	1.91 	^{+	0.40 	}_{	-0.40 	})\times
10^{	53	}$ & $(	2.32 	^{+	0.10 	}_{	-0.09 	})\times 10^{	53	} $ & 7.6 \\
080607	&	3.036	& $	-1.08	^{+	0.07	}_{	-0.06	}$ & $	-2.25
^{+	fix	}_{	-fix	}$ & $	419.0 	^{+	46.0 	}_{	-38.0 	}$ & $(	1.67
^{+	0.33 	}_{	-0.33 	})\times 10^{	-5	}$ & $(	8.57 	^{+	0.50 	}_{
-0.45 	})\times 10^{	-5	}$ & $(	1.29 	^{+	0.26 	}_{	-0.26 	})\times
10^{	54	}$ & $(	1.64 	^{+	0.10 	}_{	-0.09 	})\times 10^{	54	} $ & 4.0 \\
080721	&	2.602	& $	-0.933	^{+	0.106	}_{	-0.084	}$ & $	-2.43
^{+	0.24	}_{	-0.42	}$ & $	485.0 	^{+	67.0 	}_{	-59.0 	}$ & $(	1.36 	^{+
0.21 	}_{	-0.21 	})\times 10^{	-5	}$ & $(	7.90 	^{+	0.55 	}_{	-0.54
})\times 10^{	-5	}$ & $(	7.24 	^{+	1.11 	}_{	-1.13 	})\times 10^{	53
}$ & $(	1.17 	^{+	0.08 	}_{	-0.08 	})\times 10^{	54	} $ & 8.3 \\
080810	&	3.35	& $	-0.91	^{+	0.12	}_{	-0.12	}$ & $	-2.25	^{+	fix
}_{	-fix	}$ & $	313.5 	^{+	73.6 	}_{	-73.6 	}$ & $(	9.76 	^{+	0.84 	}_{
-0.84 	})\times 10^{	-7	}$ & $(	1.80 	^{+	0.13 	}_{	-0.13 	})\times
10^{	-5	}$ & $(	9.56 	^{+	0.83 	}_{	-0.83 	})\times 10^{	52	}$ & $(
4.05 	^{+	0.29 	}_{	-0.29 	})\times 10^{	53	} $ & 24.4 \\
080913$^{\ddagger}$$^{(\sharp)}$	&	6.695	& $	-0.46	^{+	0.7	}_{	-0.7	}$ & $	-2.25	^{+	fix
}_{	-fix	}$ & $	93.1 	^{+	56.1 	}_{	-56.1 	}$ & $(	2.31 	^{+	0.33 	}_{
-0.46 	})\times 10^{	-7	}$ & $(	1.15 	^{+	0.12 	}_{	-0.12 	})\times
10^{	-6	}$ & $(	1.15 	^{+	0.16 	}_{	-0.23 	})\times 10^{	53	}$ & $(
7.44 	^{+	0.80 	}_{	-0.80 	})\times 10^{	52	} $ & 1.0 \\
080916A	&	0.689	& $	-0.9	^{+	0.1	}_{	-0.1	}$ & $	-2.25	^{+	fix	}_{
-fix	}$ & $	109.0 	^{+	9.0 	}_{	-9.0 	}$ & $(	1.06 	^{+	0.16 	}_{	-0.16
})\times 10^{	-6	}$ & $(	2.19 	^{+	0.73 	}_{	-0.73 	})\times 10^{	-5
}$ & $(	2.15 	^{+	0.33 	}_{	-0.33 	})\times 10^{	51	}$ & $(	2.63
^{+	0.87 	}_{	-0.87 	})\times 10^{	52	} $ & 35.5 \\
081121 &	2.512	& $	-0.77	^{+	0.15	}_{	-0.14	}$ & $	-2.51
^{+	0.31	}_{	-0.66	}$ & $	248.0 	^{+	38.0 	}_{	-32.0 	}$ & $(	1.87 	^{+
0.50 	}_{	-0.46 	})\times 10^{	-6	}$ & $(	1.73 	^{+	0.35 	}_{	-0.29
})\times 10^{	-5	}$ & $(	9.15 	^{+	2.45 	}_{	-2.25 	})\times 10^{	52
}$ & $(	2.41 	^{+	0.49 	}_{	-0.40 	})\times 10^{	53	} $ & 5.1 \\
081222	&	2.77	& $	-0.55	^{+	0.07	}_{	-0.07	}$ & $	-2.1	^{+	0.06
}_{	-0.06	}$ & $	134.0 	^{+	9.0 	}_{	-9.0 	}$ & $(	2.62 	^{+	0.39 	}_{
-0.36 	})\times 10^{	-6	}$ & $(	1.70 	^{+	0.14 	}_{	-0.14 	})\times
10^{	-5	}$ & $(	1.63 	^{+	0.24 	}_{	-0.22 	})\times 10^{	53	}$ & $(
2.80 	^{+	0.23 	}_{	-0.22 	})\times 10^{	53	} $ & 8.0 \\
090102	&	1.547	& $	-0.86	^{+	0.14	}_{	-0.13	}$ & $	-2.25
^{+	fix	}_{	-fix	}$ & $	451.0 	^{+	73.0 	}_{	-58.0 	}$ & $(	3.90
^{+	0.57 	}_{	-0.55 	})\times 10^{	-6	}$ & $(	3.66 	^{+	0.34 	}_{
-0.34 	})\times 10^{	-5	}$ & $(	5.83 	^{+	0.84 	}_{	-0.82 	})\times
10^{	52	}$ & $(	2.15 	^{+	0.20 	}_{	-0.20 	})\times 10^{	53	} $ & 11.8 \\
090323	&	3.57	& $	-0.96	^{+	0.12	}_{	-0.09	}$ & $	-2.09	^{+
0.16	}_{	-0.22	}$ & $	416.0 	^{+	76.0 	}_{	-73.0 	}$ & $(	3.80 	^{+
0.73 	}_{	-0.64 	})\times 10^{	-6	}$ & $(	1.48 	^{+	0.22 	}_{	-0.17
})\times 10^{	-4	}$ & $(	4.34 	^{+	0.83 	}_{	-0.73 	})\times 10^{	53
}$ & $(	3.70 	^{+	0.55 	}_{	-0.43 	})\times 10^{	54	} $ & 35.0\\
090328	&	0.736	& $	-0.93	^{+	0.02	}_{	-0.02	}$ & $	-2.2	^{+	0.1
}_{	-0.1	}$ & $	653.0 	^{+	45.0 	}_{	-45.0 	}$ & $(	7.28 	^{+	0.54 	}_{
-0.49 	})\times 10^{	-6	}$ & $(	1.37 	^{+	0.08 	}_{	-0.07 	})\times
10^{	-4	}$ & $(	1.74 	^{+	0.13 	}_{	-0.12 	})\times 10^{	52	}$ & $(
1.89 	^{+	0.11 	}_{	-0.10 	})\times 10^{	53	} $ & 46.1 \\
090423$^{\ddagger}$$^{(\natural)}$	&	8.3	& $	-0.77	^{+	0.35	}_{	-0.35	}$ & $	-2.25	^{+	fix
}_{	-fix	}$ & $	82.0 	^{+	15.0 	}_{	-15.0 	}$ & $(	3.24 	^{+	0.61 	}_{
-0.75 	})\times 10^{	-7	}$ & $(	1.17 	^{+	0.32 	}_{	-0.32 	})\times
10^{	-6	}$ & $(	2.59 	^{+	0.48 	}_{	-0.59 	})\times 10^{	53	}$ & $(
1.01 	^{+	0.28 	}_{	-0.28 	})\times 10^{	53	} $ & 1.3 \\
090424	&	0.544	& $	-0.9	^{+	0.02	}_{	-0.02	}$ & $	-2.9	^{+	0.1
}_{	-0.1	}$ & $	177.0 	^{+	3.0 	}_{	-3.0 	}$ & $(	1.80 	^{+	0.12 	}_{
-0.11 	})\times 10^{	-5	}$ & $(	5.85 	^{+	0.22 	}_{	-0.20 	})\times
10^{	-5	}$ & $(	2.07 	^{+	0.13 	}_{	-0.12 	})\times 10^{	52	}$ & $(
4.35 	^{+	0.16 	}_{	-0.15 	})\times 10^{	52	} $ & 33.7 \\
090516  & 4.109 & $-1.51^{+0.11}_{-0.11} $ & $-2.25^{+fix}_{fix} $ & $185.6^{+98.4}_{-42.5} $ & $(5.93^{+0.24}_{-0.44}) \times 10^{-7} $ & $(2.95^{+0.64}_{-0.64}) \times 10^{-5} $ & $9.51^{+0.39}_{-0.70}) \times 10^{52} $ & $9.26^{+2.02}_{-2.02}) \times 10^{53} $ & 58.7 \\
090618  & 0.54  & $-1.26^{+0.06}_{-0.02} $ & $-2.5^{+0.15}_{-0.33} $ & $155.5^{+11.1}_{-10.5} $ & $(8.58^{+0.85}_{-1.07}) \times 10^{-6} $ & $(3.39^{+0.23}_{-0.28}) \times 10^{-4} $ & $9.63^{+0.95}_{-1.20}) \times 10^{51} $ & $2.47^{+0.17}_{-0.20}) \times 10^{53} $ & 100.6\\
090715B  & 3  & $-1.1^{+0.4}_{-0.34} $ & $-2.25^{+fix}_{fix} $ & $134^{+56}_{-30} $ & $(8.96^{+2.49}_{-2.49}) \times 10^{-7} $ & $(1.09^{+0.18}_{-0.13}) \times 10^{-5} $ & $6.73^{+1.87}_{-1.87}) \times 10^{52} $ & $2.05^{+0.33}_{-0.24}) \times 10^{53} $ & 25.0 \\
090812  & 2.452  & $-1.03^{+0.07}_{-0.07} $ & $-2.5^{+fix}_{fix} $ & $572^{+251}_{-159} $ & $(2.17^{+0.19}_{-0.28}) \times 10^{-6} $ & $(3.15^{+0.41}_{-0.41}) \times 10^{-5} $ & $1.00^{+0.09}_{-0.13}) \times 10^{53} $ & $4.21^{+0.55}_{-0.55}) \times 10^{53} $ & 20.3\\
090902B  & 1.822  & $-0.696^{+0.012}_{-0.012} $ & $-3.85^{+0.21}_{-0.31} $ & $775^{+11}_{-11} $ & $(2.72^{+0.03}_{-0.03}) \times 10^{-5} $ & $(3.78^{+0.03}_{-0.03}) \times 10^{-4} $ & $6.09^{+0.06}_{-0.07}) \times 10^{53} $ & $3.00^{+0.02}_{-0.02}) \times 10^{54} $ & 8.9 \\
090926  & 2.1062  & $-0.75^{+0.01}_{-0.01} $ & $-2.59^{+0.04}_{-0.05} $ & $314^{+4}_{-4} $ & $(1.82^{+0.03}_{-0.03}) \times 10^{-5} $ & $(1.76^{+0.50}_{-0.50}) \times 10^{-5} $ & $5.79^{+0.09}_{-0.10}) \times 10^{53} $ & $1.80^{+0.51}_{-0.51}) \times 10^{53} $ & 6.7 \\
090926B  & 1.24  & $-0.52^{+0.24}_{-0.24} $ & $-2.25^{+fix}_{fix} $ & $78.3^{+7}_{-7} $ & $(5.55^{+0.74}_{-0.83}) \times 10^{-7} $ & $(1.66^{+0.05}_{-0.05}) \times 10^{-5} $ & $4.79^{+0.64}_{-0.72}) \times 10^{51} $ & $6.39^{+0.18}_{-0.18}) \times 10^{52} $ & 8.9 \\
091018  & 0.971  & $-1.77^{+0.24}_{-0.24} $ & $-2.25^{+fix}_{fix} $ & $19.2^{+18}_{-11} $ & $(1.82^{+0.24}_{-1.44}) \times 10^{-6} $ & $(4.33^{+0.93}_{-0.58}) \times 10^{-6} $ & $8.61^{+1.13}_{-6.81}) \times 10^{51} $ & $1.04^{+0.22}_{-0.14}) \times 10^{52} $ & 5.1 \\
091020  & 1.71  & $-0.2^{+0.4}_{-0.4} $ & $-1.7^{+0.02}_{-0.02} $ & $47.9^{+7.1}_{-7.1} $ & $(1.63^{+0.18}_{-0.22}) \times 10^{-6} $ & $(1.68^{+0.37}_{-0.36}) \times 10^{-5} $ & $3.11^{+0.35}_{-0.42}) \times 10^{52} $ & $1.18^{+0.26}_{-0.26}) \times 10^{53} $ & 13.7 \\
091029  & 2.752  & $-1.46^{+0.27}_{-0.27} $ & $-2.25^{+fix}_{fix} $ & $61.4^{+17.5}_{-17.5} $ & $(2.82^{+0.16}_{-0.16}) \times 10^{-7} $ & $(5.84^{+0.53}_{-0.30}) \times 10^{-6} $ & $1.72^{+0.10}_{-0.10}) \times 10^{52} $ & $9.50^{+0.86}_{-0.48}) \times 10^{52} $ & 13.3 \\
091127  & 0.49  & $-1.27^{+0.06}_{-0.06} $ & $-2.2^{+0.02}_{-0.02} $ & $36^{+2}_{-2} $ & $(4.03^{+0.17}_{-0.17}) \times 10^{-6} $ & $(2.65^{+0.05}_{-0.05}) \times 10^{-5} $ & $3.58^{+0.15}_{-0.15}) \times 10^{51} $ & $1.58^{+0.03}_{-0.03}) \times 10^{52} $ & 6.0 \\
091208B  & 1.063  & $-1.44^{+0.07}_{-0.06} $ & $-2.32^{+0.19}_{-0.47} $ & $124^{+20.1}_{-19.4} $ & $(3.47^{+0.52}_{-0.69}) \times 10^{-6} $ & $(7.78^{+0.77}_{-0.99}) \times 10^{-6} $ & $2.05^{+0.31}_{-0.41}) \times 10^{52} $ & $2.23^{+0.22}_{-0.28}) \times 10^{52} $ & 7.2 \\
\end{longtable}



\begin{longtable}{lcccccccccc}
  \caption{Spectral parameters of 2 low luminosity GRBs and 
6 outliers excluded from the database of Table~\ref{tab:data1}.}
\label{tab:excludedata1}
  \hline              
    &          &          &         & $E^{obs}_p$ & $F_{p,bol}$              & $S_{\rm bol}$   & $L_p$          & $E_{\rm iso}$ & $T^{obs}_{90}/(1+z)$\\ 
GRB & redshift & $\alpha$ & $\beta$ & (keV)       & (erg~cm$^{-2}$~s$^{-1}$) & (erg~cm$^{-2}$) & (erg~s$^{-1}$) & (erg)         & (sec)\\  
\endfirsthead
  \hline
    &          &          &         & $E^{obs}_p$ & $F_{p,bol}$              & $S_{\rm bol}$   & $L_p$          & $E_{\rm iso}$ & $T^{obs}_{90}/(1+z)$\\ 
GRB & redshift & $\alpha$ & $\beta$ & (keV)       & (erg~cm$^{-2}$~s$^{-1}$) & (erg~cm$^{-2}$) & (erg~s$^{-1}$) & (erg)         & (sec)\\ 
\endhead
  \hline
\endfoot
  \hline
\multicolumn{11}{l}{$\dagger$~Low luminosity GRB.}\\
\multicolumn{11}{l}{$\ddagger$~Outlier which locates beyond 3~$\sigma$ from the best fit function of the $E_{p}$--$L_{p}$ and/or the $E_{p}$--$E_{\rm iso}$ relations.}\\
\endlastfoot
  \hline
980425$^{\dagger}$	&	0.0085	& $	-0.97	^{+	0.16	}_{	-0.16	}$ & $	-2.06	^{+	0.09	}_{
-0.09	}$ & $	54.9 	^{+	11.5 	}_{	-11.5 	}$ & $(	5.78	^{+	1.01 	}_{
-0.95 	})\times 10^{	-7	}$ & $(	5.97 	^{+	1.18 	}_{	-1.19 	})\times
10^{	-6	}$ & $(	1.03 	^{+	0.18 	}_{	-0.17 	})\times 10^{	47	}$ & $(
1.04 	^{+	0.21 	}_{	-0.21 	})\times 10^{	48	} $ & 39.7 \\
050223$^{\ddagger}$  & 0.5915  & $-1.5^{+0.42}_{-0.42} $ & $-2.25^{+fix}_{fix} $ & $69.1^{+34}_{-34} $ & $(1.18^{+0.17}_{-0.17}) \times 10^{-7} $ & $(1.91^{+0.19}_{-0.19}) \times 10^{-6} $ & $(1.67^{+0.24}_{-0.24}) \times 10^{50} $ & $(1.69^{+0.17}_{-0.17}) \times 10^{51}$ & 14.5 \\
050803$^{\ddagger}$  & 0.422  & $-0.99^{+0.37}_{-0.37} $ & $-2.25^{+fix}_{fix} $ & $97^{+34}_{-34} $ & $(2.59^{+0.35}_{-0.41}) \times 10^{-7} $ & $(6.99^{+0.54}_{-0.54}) \times 10^{-6} $ & $(1.61^{+0.22}_{-0.26}) \times 10^{50} $ & $(3.06^{+0.24}_{-0.24}) \times 10^{51}$ & 77.4\\
050904$^{\ddagger}$	&	6.295	& $	-1.11	^{+	0.11	}_{	-0.11	}$ & $	-2.25	^{+	fix	}_{
-fix	}$ & $	436 	^{+	335 	}_{	-151 	}$ & $(	1.95	^{+	0.53 	}_{
-0.55 	})\times 10^{	-7	}$ & $(	1.77 	^{+	0.21 	}_{	-0.57 	})\times
10^{	-5	}$ & $(	8.48 	^{+	2.32 	}_{	-2.38 	})\times 10^{	52	}$ & $(
1.05 	^{+	0.13 	}_{	-0.34 	})\times 10^{	54	} $ & 30.8 \\
060218$^{\dagger}$	&	0.0331	& $	-1.00	^{+	fix	}_{	-fix	}$ & $	-2.50	^{+	0.10	}_{
-0.10	}$ & $	4.9 	^{+	0.3 	}_{	-0.3 	}$ & $(	1.65	^{+	0.59 	}_{
-0.64 	})\times 10^{	-8	}$ & $(	3.72 	^{+	0.29 	}_{	-0.30 	})\times
10^{	-6	}$ & $(	4.27 	^{+	1.54 	}_{	-1.66 	})\times 10^{	46	}$ & $(
9.32 	^{+	0.73 	}_{	-0.76 	})\times 10^{	48	} $ & --- \\
070714B$^{\ddagger}$	&	0.92	& $	-0.86	^{+	0.1	}_{	-0.1	}$ & $	-2.25	^{+
fix	}_{	-fix	}$ & $	1120.0 	^{+	780.0 	}_{	-380.0 	}$ & $(	3.53 	^{+
0.42 	}_{	-0.91 	})\times 10^{	-6	}$ & $(	5.07 	^{+	1.78 	}_{	-0.82
})\times 10^{	-6	}$ & $(	1.45 	^{+	0.17 	}_{	-0.38 	})\times 10^{	52
}$ & $(	1.09 	^{+	0.38 	}_{	-0.18 	})\times 10^{	52	} $ & --- \\
090418$^{\ddagger}$	&	1.608	& $	-1.3	^{+	0.09	}_{	-0.09	}$ & $	-2.25
^{+	fix	}_{	-fix	}$ & $	610.0 	^{+	530.0 	}_{	-164.0 	}$ & $(	7.18
^{+	1.17 	}_{	-2.06 	})\times 10^{	-7	}$ & $(	2.53 	^{+	0.30 	}_{
-0.30 	})\times 10^{	-5	}$ & $(	1.18 	^{+	0.19 	}_{	-0.34 	})\times
10^{	52	}$ & $(	1.60 	^{+	0.19 	}_{	-0.19 	})\times 10^{	53	} $ & 21.5 \\
091003$^{\ddagger}$  & 0.8969  & $-1.13^{+0.01}_{-0.01} $ & $-2.64^{+0.24}_{-0.24} $ & $486.2^{+23.6}_{-23.6} $ & $(7.15^{+0.50}_{-0.39}) \times 10^{-6} $ & $(5.05^{+0.84}_{-0.73}) \times 10^{-6} $ & $(2.78^{+0.20}_{-0.15}) \times 10^{52} $ & $(1.04^{+0.17}_{-0.15}) \times 10^{52} $ & 11.1 \\
\end{longtable}


\begin{longtable}{lcccccccccc}
  \caption{Spectral parameters of GRBs which have some ambiguities 
in the redshift or spectral parameters. 
These samples are excluded from the database of Table~\ref{tab:data1}.}
\label{tab:excludedata2}
  \hline              
    &          &          &         & $E^{obs}_p$ & $F_{p,bol}$              & $S_{\rm bol}$   & $L_p$          & $E_{\rm iso}$ & $T^{obs}_{90}/(1+z)$\\ 
GRB & redshift & $\alpha$ & $\beta$ & (keV)       & (erg~cm$^{-2}$~s$^{-1}$) & (erg~cm$^{-2}$) & (erg~s$^{-1}$) & (erg)         & (sec)\\  
\endfirsthead
  \hline
    &          &          &         & $E^{obs}_p$ & $F_{p,bol}$              & $S_{\rm bol}$   & $L_p$          & $E_{\rm iso}$ & $T^{obs}_{90}/(1+z)$\\ 
GRB & redshift & $\alpha$ & $\beta$ & (keV)       & (erg~cm$^{-2}$~s$^{-1}$) & (erg~cm$^{-2}$) & (erg~s$^{-1}$) & (erg)         & (sec)\\ 
\endhead
  \hline
\endfoot
  \hline
\multicolumn{11}{l}{$*$~Redshift ambiguity is large.}\\
\multicolumn{11}{l}{$\sharp$~Can not determine the $E^{obs}_{p}$ value 
because of $\beta > -2.0$.}\\
\multicolumn{11}{l}{$\flat$~Ambiguity of spectral parameters are 
rather large.}\\
\multicolumn{11}{l}{$\natural$~Short GRBs whose peak flux is measured in 
millisecond time scale and we can not convert them into 1~second peak flux.}\\
\endlastfoot
  \hline
980326$^{*}$	&	0.9--1.1	& $	-0.93	^{+	0.09	}_{	-0.08	}$ & $	-2.96	^{+	0.21	}_{
-0.51	}$ & $	35.5 	^{+	18.0 	}_{	-18.0 	}$ & $(	9.05	^{+	4.14 	}_{
-2.82 	})\times 10^{	-7	}$ & $(	1.26 	^{+	0.54 	}_{	-0.54 	})\times
10^{	-6	}$ & (2.45--8.47)$\times 10^{51}$ & (1.49--5.50)$\times 10^{51}$ & 4.3--4.7\\
980329$^{*}$	&	2.0--3.9	& $	-0.79	^{+	0.03	}_{	-0.03	}$ & $	-2.27	^{+	0.04	}_{
-0.04	}$ & $	249.2 	^{+	32.8 	}_{	-43.2 	}$ & $(	5.89	^{+	0.33 	}_{
-0.34 	})\times 10^{	-6	}$ & $(	8.95 	^{+	0.11 	}_{	-0.11 	})\times
10^{	-5	}$ & (1.56--8.78)$\times 10^{53}$ & (8.28--26.1)$\times 10^{53}$ & 3.8--6.2\\
980703$^{\sharp}$	&	0.966	& $	-0.80	^{+	0.22	}_{	-0.16	}$ & $	-1.60	^{+	0.06	}_{
-0.09	}$ & $	(> 76.3) $ & $(	2.56	^{+	0.28 	}_{
-0.49 	})\times 10^{	-6	}$ & $(	9.96 	^{+	1.02 	}_{	-1.14 	})\times
10^{	-5	}$ & $(	1.20 	^{+	0.13 	}_{	-0.23 	})\times 10^{	52	}$ & $(
2.37 	^{+	0.24 	}_{	-0.27 	})\times 10^{	53	} $ & 209.4 \\
000214$^{*}$	&	0.37--0.47	& $	-1.62	^{+	0.13	}_{	-0.13	}$ & $	-2.10	^{+	fix	}_{
-fix	}$ & $	(> 82) $ & $(	9.84	^{+	0.49 	}_{
-0.49 	})\times 10^{	-6	}$ & $(	3.49 	^{+	0.14 	}_{	-0.10 	})\times
10^{	-5	}$ & (4.30--8.35)$\times 10^{51} $ & (1.14--2.00)$\times 10^{52}$ & 6.8--7.3\\
010222$^{\sharp}$	&	1.437	& $	-1.35	^{+	0.19	}_{	-0.19	}$ & $	-1.64	^{+	0.02	}_{
-0.02	}$ & $	(> 358) $ & $(	2.30	^{+	0.10 	}_{
-0.10 	})\times 10^{	-5	}$ & $(	2.47 	^{+	0.13 	}_{	-0.12 	})\times
10^{	-4	}$ & $(	2.87 	^{+	0.13 	}_{	-0.12 	})\times 10^{	53	}$ & $(
1.26 	^{+	0.06 	}_{	-0.06 	})\times 10^{	54	} $ & 53.3 \\
050709$^{\natural}$	&	0.1606	& $	-0.53	^{+	0.12	}_{	-0.13	}$ & $	-2.25	^{+	fix	}_{
  -fix}$ & $  83.9  ^{+ 11}_{-8.3}$ & $(	8.54	^{+	1.07 	}_{
-1.12 	})\times 10^{	-6	}$ & $(	7.23 	^{+	0.69 	}_{	-0.75 	})\times
10^{	-6	}$ & $(	5.88 	^{+	0.74	}_{	- 0.77 	})\times 10^{	50	}$ & $(
4.29 	^{+	0.41 	}_{   - 0.45 	})\times 10^{	50	} $ & 0.07 \\
050824$^{\flat}$	&	0.83	& $	-1.00	^{+	fix	}_{	-fix	}$ & $	-2.35	^{+	0.88	}_{
-0.48	}$ & $	(<12.7) $ & $	(< 7.25 \times 10^{	-8	})$ & $(	< 7.18 \times
10^{	-7	}) $ & $(	< 2.33 \times 10^{	50	})$ & $(
< 1.26 \times 10^{	51	}) $ & 13.7 \\
051221$^{\natural}$	&	0.5465	& $	-1.08	^{+	0.13	}_{	-0.14	}$ & $	-2.25	^{+	fix	}_{
  -fix}$ & $  402  ^{+ 93}_{-72}$ & $(	5.69	^{+	0.25 	}_{
-3.10 	})\times 10^{	-5	}$ & $(	3.96 	^{+	0.12 	}_{	-2.11 	})\times
10^{	-6	}$ & $(	6.59 	^{+	0.29 	}_{	- 3.59 	})\times 10^{	52	}$ & $(
2.96 	^{+	0.09 	}_{   - 0.16 	})\times 10^{	51	} $ & 0.128 \\
060418$^{\flat}$	&	1.489	& $	-1.66	^{+	0.05	}_{	-0.05	}$ & $	-2.25	^{+	fix	}_{
-fix	}$ & $	230 	^{+	fix 	}_{	-fix 	}$ & $(	1.40	^{+	0.08 	}_{
-0.11 	})\times 10^{	-6	}$ & $(	2.61 	^{+	0.08 	}_{	-0.08 	})\times
10^{	-5	}$ & $(	1.90 	^{+	0.11 	}_{	-0.15 	})\times 10^{	52	}$ & $(
1.43 	^{+	0.04 	}_{	-0.04 	})\times 10^{	53	} $ & 20.9 \\
060614$^{\flat}$	&	0.125	& $	-1.00	^{+	fix	}_{	-fix	}$ & $	-2.14	^{+	0.04	}_{
-0.04	}$ & $	55 	^{+	45 	}_{	-45 	}$ & $(	2.03	^{+	0.32 	}_{
-0.24 	})\times 10^{	-6	}$ & $(	5.41 	^{+	0.41 	}_{	-0.37 	})\times
10^{	-5	}$ & $(	8.08 	^{+	1.29 	}_{	-0.94 	})\times 10^{	49	}$ & $(
1.91 	^{+	0.14 	}_{	-0.13 	})\times 10^{	51	} $ & 90.7 \\
061006$^{\natural}$	&	0.4377	& $	-0.62	^{+	0.18	}_{	-0.21	}$ & $	-2.25	^{+	fix	}_{
  -fix}$ & $  664  ^{+ 227}_{-114}$ & $(	2.89	^{+	0.56 	}_{
-1.64 	})\times 10^{	-5	}$ & $(	4.84 	^{+	0.42 	}_{	-2.63 	})\times
10^{	-6	}$ & $(	1.97 	^{+	0.38 	}_{	- 1.12 	})\times 10^{	52	}$ & $(
2.29 	^{+	0.20 	}_{   - 0.13 	})\times 10^{	51	} $ & 0.42 \\
070714$^{\natural}$	&	0.92	& $	-0.86	^{+	0.10	}_{	-0.10	}$ & $	-2.25	^{+	fix	}_{
  -fix}$ & $  1120 ^{+ 780}_{-380}$ & $(	3.50	^{+	0.42 	}_{
-0.91 	})\times 10^{	-6	}$ & $(	5.03 	^{+	1.77 	}_{	-0.82 	})\times
10^{	-6	}$ & $(	1.44 	^{+	0.17 	}_{	- 0.38 	})\times 10^{	52	}$ & $(
1.08 	^{+	0.38 	}_{   - 0.18 	})\times 10^{	52	} $ & 2.0 \\
080319C$^{\sharp}$	&	1.95	& $	-1.01	^{+	0.13	}_{	-0.13	}$ & $	-1.87	^{+	0.15	}_{
-0.53}$ & $	(> 307)  $ & $(	2.86	^{+	0.68 	}_{
-0.58 	})\times 10^{	-6	}$ & $(	1.48 	^{+	0.34 	}_{	-0.20 	})\times
10^{	-5	}$ & $(	7.60 	^{+	1.81 	}_{	-1.55 	})\times 10^{	52	}$ & $(
1.33 	^{+	0.31 	}_{	-0.18 	})\times 10^{	53	} $ & 11.5 \\
\end{longtable}
}


%

\end{document}